\documentclass[12pt]{iopart}
\usepackage{graphicx}
\usepackage{xcolor} % Highlighting
\makeatletter
\expandafter\let\csname equation*\endcsname\relax
\expandafter\let\csname endequation*\endcsname\relax
\makeatother
\usepackage{amsmath}
\usepackage{mathtools}
\usepackage{amssymb}
\usepackage{caption}
\usepackage{subcaption}
\usepackage{hyperref}
\usepackage{threeparttable}
\usepackage{float} % for the H specifier
\usepackage[verbose]{placeins}
\usepackage{booktabs}
\usepackage[top=0.8in, bottom=1in,]{geometry}
\begin{document}

% \title[]{An Analytical Framework for Quantifying Battery Performance: Solid Solution Electrodes}
\title[]{Scaling and Analytical Approximation of Porous Electrode Theory for Reaction-limited Batteries}

\author{Shakul Pathak$^{1}$, Martin Z. Bazant$^{1,2,z}$}
\address{$^1$ Department of Chemical Engineering, Massachusetts Institute of Technology, Cambridge, Massachusetts 02139, USA}
\address{$^2$ Department of Mathematics, Massachusetts Institute of Technology, Cambridge, Massachusetts 02139, USA}
% \address{$^*$ ECS Members.}
\address{$^z$ Corresponding author.}
\ead{\mailto{bazant@mit.edu} (M.Z.B.)}

\vspace{10pt}
\begin{abstract}
Porous electrode theory (PET) provides essential insights into electrochemical states, but its computational complexity hinders real-time control and obscures scaling relations. To bridge the gap between high-fidelity simulations and reduced-order models, we present a framework of scaling analysis and analytical approximations. By assuming high-performance electrodes minimize transport limitations and overpotentials, we derive a simplified ``lean model'' governed by four dimensionless numbers: (i) a traditional Damköhler number, $Da$, scaling the characteristic reaction
rate to the diffusion rate in the electrolyte-filled
pores; (ii) the “process Damköhler number,” $Da_p$, scaling the reaction rate to
the applied capacity utilization rate (C-rate); (iii) the “wiring Damköhler
number,” $Da_w$, scaling the reaction rate to an effective electromigration rate for
ions in the pores in series with electrons in the conducting matrix; and (iv) the
“capacitive Damköhler number,” $Da_c$, comparing the rates of Faradaic reactions
and double-layer charging. For batteries, we derive analytical solutions for standard protocols, including galvanostatic discharge, chronoamperometry, and electrochemical impedance spectroscopy. Validated against numerical simulations of a practical NMC half-cell, our formulae show excellent agreement at negligible computational cost. This interpretable, physics-based framework accelerates battery design and state estimation while unifying the modeling of batteries, supercapacitors, fuel cells, and other porous electrode systems.
\end{abstract}
\vspace{2pc}
\maketitle
\section{Introduction}
\label{sec:Intro}
Porous electrode theory (PET) provides a unifying and widely accepted mathematical framework for physics-based modeling of electrochemical systems~\cite{newman2021electrochemical}, such as batteries, supercapacitors, fuel cells, electrolyzers, and desalination and electrosorption systems, which maximize capacity and power density through the use of porous electrodes. Mathematical models based on PET are used to support research, scale-up, and deployment of lithium-ion batteries (LIBs) \cite{lyu2025next}, driven by growing demand for  electric vehicles, portable electronics, robots, and grid-scale energy storage \cite{zubi2018lithium}. Mathematical modeling can also play a crucial in the engineering of battery-like systems for selective ion extraction~\cite{wu2022lithium}, as well as in in electro-sorption based capacitive deionization, energy storage and energy harvesting~\cite{alkhadra2022electrochemical}.

Here we focus on the important case of a Li-ion battery constructed from two porous electrodes---a cathode (e.g., $\mathrm{Li}_{x}\mathrm{Ni}_{y}\mathrm{Mn}_{z}\mathrm{Co}_{w}\mathrm{O}_{2}$ (NMC), $\mathrm{Li}_{x}\mathrm{CoO}_{2}$ (LCO)) and an anode (e.g., graphite, $\mathrm{Li}_{4+x}\mathrm{Ti}_{5}\mathrm{O}_{12}$ (LTO))---separated by a porous membrane. The battery's response is governed by a complex interplay of physical phenomena occurring across multiple scales within the electrode's composite matrix.
This matrix consists of active material, conductive additives, binder, and electrolyte \cite{newman2021electrochemical}.
The complexity of these interactions makes predicting battery response difficult.
Furthermore, interpreting datasets from experimental characterization including electrochemical tests (e.g., galvanostatic cycling, impedance spectroscopy) and materials analysis (e.g., X-ray diffraction, electron microscopy) is challenging and often requires sophisticated models \cite{hu2025integrating}.

A spectrum of battery models is available today, ranging in physical fidelity and computational cost \cite{ramadesigan2012modeling,brosa2022continuum}.
At one end, equivalent circuit models (ECMs) are widely applied in battery management systems (BMS) for state-of-charge estimation due to their computational efficiency \cite{lin2019modeling, zheng2024review}.
However, their empirical nature limits their predictive power, particularly as the cell degrades and develops internal heterogeneities, leading to physically inconsistent predictions \cite{chen2022porous}.
At the other end of the spectrum are multiphysics models founded upon PET, as introduced by Newman and collaborators \cite{newman1975porous}, building on de Levie's transmission-line model of porous-electrode ``wiring" and distributed capacitance~\cite{de1964porous}. 
Over the past 30 years, the Doyle--Fuller--Newman (DFN) model \cite{doyle1993modeling} and its variants have provided the standard macroscopic description of battery behavior rooted in microscopic transport and kinetics. These models are widely employed for cell design and performance prediction~\cite{ramadesigan2012modeling}, although they have some limitations. 
In particular, the classical DFN framework is best suited for solid-solution active materials and inherently 
cannot  describe common phase-separating materials, such as lithium iron phosphate (LFP) or graphite, where the cell voltage is an emergent property of inhomogeneous concentration profiles \cite{ferguson2012nonequilibrium}.
To address this limitation, Multiphase Porous Electrode Theory (MPET) was developed \cite{smith2017multiphase}, which generalizes the classical framework using nonequilibrium thermodynamics~\cite{bazant2013theory} and Cahn-Hilliard phase-field models \cite{cahn1958free} to describe a wider range of active materials \cite{guo2016li}.
This approach successfully captures complex behaviors, such as mosaic instabilities~\cite{ferguson2014phase,li2014current} and electro-autocatalysis~\cite{bazant2017thermodynamic,park2021fictitious}, and has been successful in predicting experimental data from phase-separating porous electrodes~\cite{zhao2023learning,guo2016li,thomas2017situ,lian2024modeling}.

While physics-based models such as DFN and MPET provide quantitative insights into battery operation, their complexity often limits their use in applications requiring real-time feedback, such as state estimation within a BMS or large-scale parameter optimization \cite{ moura2016battery, wang2022review, zhou2025learning}.
Volume-averaged approaches such as MPET and DFN perform numerical multiscale simulations of the internal battery state but rely on assumptions about system size, heterogeneity, and transport mechanisms.
This creates a bottleneck in model fidelity, often leading to qualitative rather than quantitative agreement in commercial systems.
Furthermore, simulations of ensembles of particles (often thousands) with more than 20 partially correlated parameters \cite{berliner2021methods} are difficult to invert given limited experimental datasets of electrode response.
Typical experimental datasets (e.g., constant current (CC), linear sweep voltammetry (LSV), voltage/current pulsing) represent the average behavior of millions of particles with diverse heterogeneities, such as size, coating density, and electrochemically active area \cite{ombrini2025modeling}.
Identifying all parameters in an MPET or DFN model is not only computationally challenging but is also an ill-posed problem requiring careful strategies to limit overfitting \cite{berliner2021nonlinear}.

The identifiability and computational speed barriers presented by numerical simulations pose significant challenges to the widespread application of physics-based approaches \cite{forman2012genetic, bizeray2018identifiability}.
This highlights the need for reduced-order models that retain the predictive power of porous electrode theory but are amenable to solutions in a reduced space of essential physical parameters.
The value of such reduced porous electrode models is already recognized.
Examples include simpler numerical models such as the single particle model (SPM) \cite{atlung1979dynamic, chaturvedi2010algorithms} or the single particle model with electrolyte (SPMe) \cite{kemper2013extended,marquis2019asymptotic, planella2021systematic}, which require fewer parameters.
The SPM considers the asymptotic limit of the DFN model under high electronic conductivity in the conductive backbone and high ionic transport in the electrolyte phase \cite{atlung1979dynamic}.
The SPMe builds beyond the SPM by making first-order corrections for ionic transport limitations in the electrolyte \cite{moura2016battery, marquis2019asymptotic}.
More sophisticated reduced-order approaches, such as population-based approaches for particles in an electrolyte bath and basis function expansions for intra-particle transport, have also been proposed \cite{zhuang2024scaling, zhao2019population}.

In this work, we perform a general scaling analysis and derive analytical approximations to model reaction-limited solid-solution LIB electrodes.
We begin from a full, coupled DFN model in the limit of fast transport in electrode particles.
We then linearize the overpotential dependence of the intercalation kinetics, guided by coupled ion-electron transfer (CIET) theory~\cite{bazant2023unified} and extensive experimental data for Li-ion batteries~\cite{zhang2025lithium}.
These approximations allow us to obtain closed-form analytical solutions for key electrochemical variables, which are validated against full numerical simulations for typical electrochemical protocols such as galvanostatic (dis)charge, chronoamperometry, and electrochemical impedance spectroscopy.

Our framework reveals specialized expressions that clarify the connection between key dimensionless groups and battery behavior.
This can greatly accelerate cell design, especially when experimental data is limited.
Furthermore, it enables high-fidelity battery state estimation via fast physics-based, data-driven, or hybrid onboard diagnostics through the integration of our model with machine learning approaches.
Our solution approach forms a basis for powerful yet simple quantitative models of industrially relevant battery materials with challenging physics, such as phase change and degradation \cite{o2022lithium, sulzer2021promise, foster2025newman}.

The rest of the paper is organized as follows: Section \ref{sec:model} formulates the governing equations, introduces assumptions for the reaction-limited regime, and derives scalings to reveal dimensionless groups that characterize battery response.
Section \ref{sec:approach} derives example solutions for key electrochemical protocols---including galvanostatic discharge, chronoamperometry, and electrochemical impedance spectroscopy.
In Section \ref{sec:results}, these solutions are benchmarked against high-fidelity numerical simulations (MPET and PyBaMM) for realistic NMC parameters.
Finally, Section \ref{sec:Discuss} summarizes the findings and discusses implications for battery state estimation.
% \section{Introduction}
% \label{sec:Intro}
% % The rapid growth of the electric vehicles market and grid-scale energy storage have intensified demand for high-performance batteries.
% Growing demand for lithium-ion batteries (LIBs) has driven significant research, scale-up and deployment that is facilitated by physics-based models \cite{lyu2025next}.
% These models are ideally able to predict battery performance, health, thermal safety, and lifetime using limited real-world data .
% Constructed from two porous electrodes---a cathode (e.g., Li$_{x}$Ni$_{y}$Mn$_{z}$Co$_{w}$O$_{2}$ (NMC), Li$_{x}$CoO$_{2}$ (LCO)) and an anode (e.g., graphite, Li$_{4+x}$Ti$_{5}$O$_{12}$ (LTO))---separated by a porous membrane, a battery's response is governed by a complex interplay of physical phenomena occurring across multiple scales within the electrode's composite matrix, which consists of active material, conductive additives, binder, and electrolyte.

\section{Porous Electrode Theory} \label{sec:model}

This section presents the governing equations and key assumptions.
With the goal of developing a reduced-order model, our work builds upon the porous-electrode theory by Newman and Tiedemann \cite{newman1975porous}, which adds electrochemical reactions and solid-state diffusion to the transmission line models of porous electrode charging introduced by de Levie~\cite{de1964porous}.
The composite electrode is modeled using volume averaged mass and charge balance equations within the electronically conducting solid phase,  the electrolyte phase and the active particles.
The resulting system of equations forms the basis for analysis in the next section.
% We assume the electrode operates in a reaction-limited regime, where the characteristic time for solid-state diffusion within an active material particle ($\tau_s = R_p^2/D_s$) is much smaller than the characteristic process timescale ($\tau_p$).
This allows for the simplification that the intercalant concentration, $c_s$, is spatially uniform within each particle.
The evolution of this volume-averaged concentration is thus governed solely by the interfacial charge-transfer current density, $j$.

\subsection{Electron Transport }

Charge conservation in the solid matrix is given by,
\begin{equation}
    -\frac{\partial i_s}{\partial x} =  -a_p j
    \label{eq:phie_simplified}
\end{equation}
$\phi_s$ is the (Galvani) potential of electrons in the solid conductive phase, $a_p$ is the active internal surface area per volume of the porous electrode, and the solid-phase current density $i_s$ is typically described by Ohm's law, $i_s = -\sigma_s \nabla \phi_s$, with an effective electronic conductivity, $\sigma_s$.

\subsection{Electrolyte Transport }

Transport in the electrolyte is described using binary concentrated solution theory \cite{newman2021electrochemical}.
Combining the anionic and cationic species conservation into one, the species and charge balance equations in the electrolyte are given by,
\begin{align}
    \epsilon_p \frac{\partial c_l}{\partial t} &= \frac{\partial}{\partial x}\left( D_{\text{eff}} \frac{\partial c_l}{\partial x}\right) -\frac{1-t_+}{F} a_p j\label{eq:cl_simplified} \\
     -\frac{\partial i_l}{\partial x} &=  a_p j \label{eq:current_simplified}
\end{align}
where $D_{\text{eff}}$ is the effective ionic diffusivity that may be a function of electrolyte composition and the ionic current density, $i_l$, is a function of gradients in electrolyte potential, $\phi_l$, and salt concentration, typically expressed in mol/L or mol/m$^3$, is $c_l$ :
\begin{equation}
    i_l = -\kappa_l \frac{\partial \phi_l}{\partial x} + \kappa_l\frac{2R_gT}{F}(1-t_+)\frac{\partial \ln c_l}{\partial x} \label{eq:il_simplified}
\end{equation}
where $\kappa_l$ is 
the electrolyte concentration dependent ionic conductivity and $R_gT/F = k_BT/e$ is the thermal voltage.
Note that $\phi_l$ is the potential with respect to a $\text{Li}/\text{Li}^+$ reference.

\subsection{ Charge Transfer Kinetics} \label{sec:ctk}

An essential aspect of PET is the model of Faradaic reaction kinetics at the internal porous electrode/electrolyte interface. For example, in this paper, we focus on the lithium ion intercalation reaction,
\begin{align}
\mathrm{Li}^+ + e^- \rightleftharpoons \mathrm{Li}_{\text{(int)}}
\end{align}
where $\mathrm{Li}_{\text{(int)}}$ is the reduced state of the reaction, consisting of an intercalated $\mathrm{Li}^+$ ion plus a nearby electron, which reduced the electrode matrix~\cite{bazant2013theory}. 
The Faradaic current density $j$ generally depends on the overpotential, $\eta = (\mu_{\mathrm{Li}_{\text{(int)}}} - \mu_{\mathrm{Li}^+} - \mu_e)/e$, defined in terms of electrochemical potentials ($\mu_i$) of the reacting species, as the free energy of reaction per charge transferred~\cite{bazant2013theory}. Various models of the reaction mechanism can be used to derive the functional form of the reaction rate depending on overpotential, temperature, and concentrations. For the concentrated electrolytes,
Frumkin effects of interfacial charge in the electric double layer~\cite{biesheuvel2009imposed} are usually neglected  due to strong charge screening, which corresponds to the Helmholtz limit of PET \cite{biesheuvel2011diffuse}. 

The standard phenomenological model of electrochemical reaction kinetics is the Butler-Volmer equation \cite{butler1932mechanism, erdey1931frage}, 
\begin{align}
    j_{BV} = k_{BV} f_{BV} (\tilde{c}_s,\tilde{c}_l)(\exp(-\alpha \tilde{\eta})-\exp((1-\alpha)\tilde{\eta}))
\end{align}
where  $k_{BV}$ is an empirical pre-factor and $f_{BV}(\tilde{c}_s,\tilde{c}_l)$ captures the concentration (or more generally, activity) dependence in the exchange current density.  
Although widely used in PET, departures from BV kinetics are often seen in the form of curved Tafel plots, especially for Li-ion batteries \cite{bai2014charge,zhang2025lithium}. To account for this, the BV equation is typically modified by adding a fitted film resistance that bends the Tafel plots \cite{scanlan2025equations}. This is a simple approach that often succeeds in fitting the curved Tafel plots but leaves open questions about the applicability of the framework beyond the fitting data \cite{morasch2023li}.
Moreover, consistent departures from BV in kinetic measurements across different materials, solid concentrations ($\tilde{c}_s$) and temperatures remain difficult to justify with a fitted film resistance.

A more systematic, physics-based approach is provided by the quantum-mechanical theory of coupled ion-electron transfer (CIET)~\cite{bazant2013theory,fraggedakis2021theory}. The BV equation can be derived from CIET theory in the limit of ``ion-coupled electron transfer" (ICET), 
\begin{align}
    j_{ICET} &=  k_{ICET} f_{ICET}(\tilde{c}_s,\tilde{c}_l)(\exp(-\alpha \tilde{\eta})- \exp((1-\alpha)\tilde{\eta})) 
\end{align}
where the free energy of ion transfer exceeds the Marcus reorganization energy, resulting in theoretical expression for the exchange current density and rate prefactor related to microscopic interfacial properties~\cite{bazant2013theory}. However, most common battery materials have  been found to exhibit CIET kinetics which are better approximated by the opposite limit of ``electron-coupled ion transfer"(ECIT) in which electron transfer is rate limiting~\cite{bai2014charge,zhang2025lithium},
\begin{align}
    j_{ECIT} &= k_{ECIT} \left(\frac{\tilde{c}_s}{1+\exp(-\tilde{\eta}_f)}-\frac{\tilde{c_l}}{1+\exp(\tilde{\eta}_f)}\right)\text{erfc}\left(\frac{\tilde{\lambda}+\sqrt{1+\sqrt{\tilde{\lambda}}+\tilde{\eta}_f^2}}{2\sqrt{\tilde{\lambda}}}\right)  
\end{align}
where $\tilde{c}_s$ and $\tilde{c}_l$ are scaled solid filling fraction and electrolyte concentration respectively, $\tilde{\eta}_f$ is the formal overpotential scaled by $R_gT/F=k_BT/e$, $\tilde{\lambda}$ is the reorganization energy scaled by thermal voltage and $k_{ECIT}$ and $k_{ICET}$, are lumped product pre-factors with terms having Arrhenius dependence on the activation energy and the coupling between the donor and acceptor states (chemisorption function)\cite{bazant2023unified}. A uniformly valid approximation interpolating between the ICET and ECIT limits is also available~\cite{bazant2013theory}.

\begin{figure}[H]
    \centering
\includegraphics[width=0.75\textwidth]{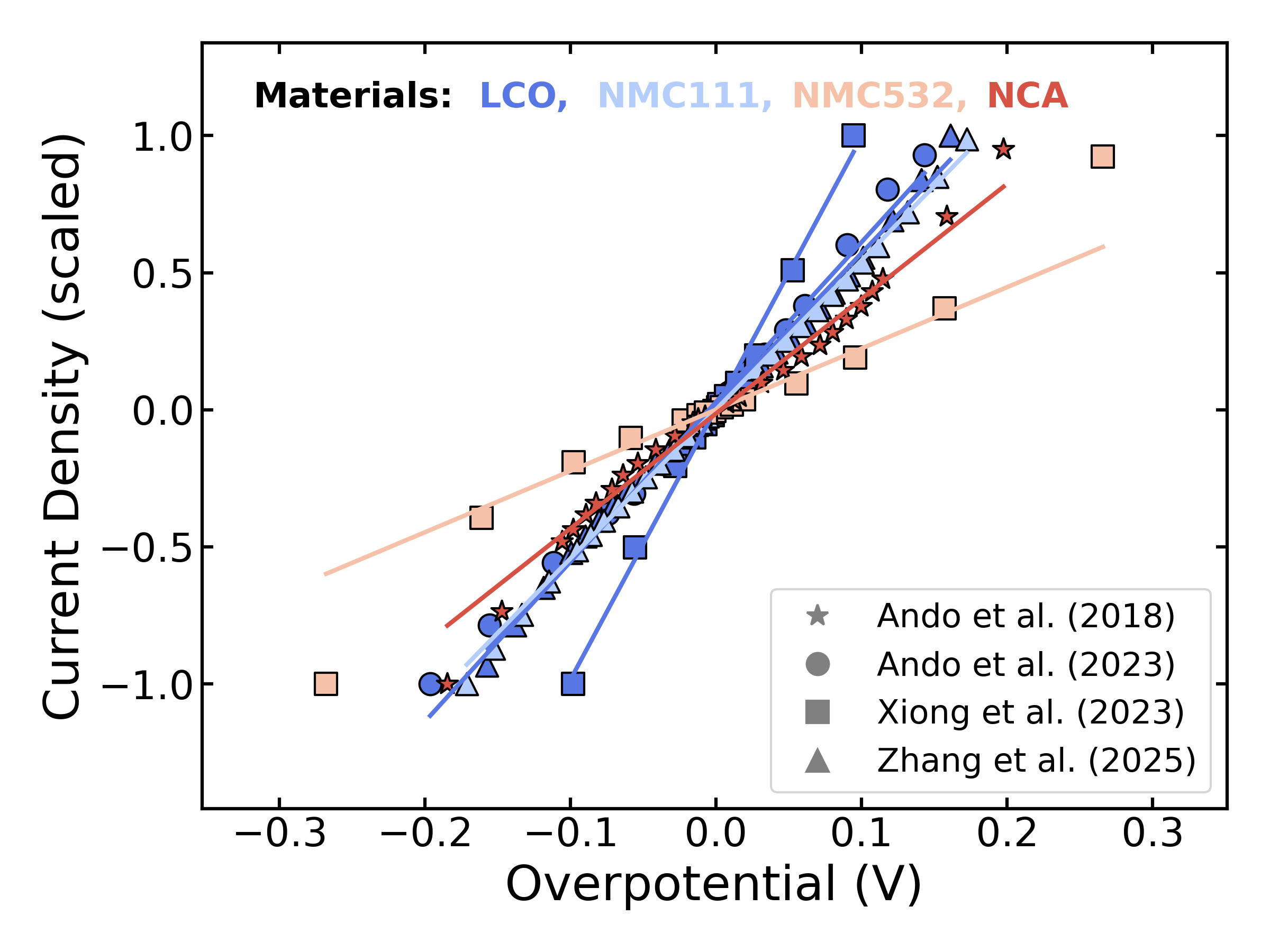}
    \vspace{-0.2cm}
    \caption{Experimental measurements of intercalation kinetics in LIBs (scatter) with linear fits (solid).
Data taken from Ref. \cite{ando2018degradation, ando2023impact, xiong2023decoupled,zhang2025lithium}. Current densities are scaled by the maximum magnitude for each dataset.}
    \label{fig:kinfit}
\end{figure}

As discussed earlier, kinetic studies of Li-ion battery electrodes often show significant deviation from exponential rise in current with overpotential predicted by Tafel's law.
The deviation typically leads to curved Tafel plots with scaling that is slower than exponential and motivates the following question: ``\textit{How well can intercalation in LIBs be approximated using a linear function of overpotential?}'' To answer this, we compiled kinetic measurements for select cathode materials (NCM, LCO, NCA) \cite{xiong2023decoupled,ando2023impact,ando2018degradation} in Figure \ref{fig:kinfit}. Note that the plot is on a linear scale. Straight-line fits to the experimentally observed reaction kinetics reveal a visibly linear trend up to overpotentials as high as 5--10$\text{k}_\text{B}\text{T}/\text{e}$. This covers a significant range of kinetic overpotentials in battery operation. Consequently, real-world battery kinetics may 
be approximated by relationships that are linear in overpotential. This approximation forms the basis for our approach in Section \ref{sec:approach}.

\subsection{Reaction-limited Active Particles} 
When transport of the intercalant (Li) within the particle can be described using a flux $\mathbf{F}_s$, the species conservation equation within the active particle becomes,
\begin{align}
     \frac{\partial c_s}{\partial t} = -\nabla\cdot \mathbf{F}_s
     \label{eqn:particledyn}
\end{align}
where $\mathbf{F}_s$ is based on irreversible thermodynamics and $c_s$ once again is the intercalation concentration in mol/m$^3$.
This equation must be specified with an initial condition i.e. typically a homogeneous concentration profile.
Since electrode particles undergo (de)insertion with the electrolyte, the boundary condition must include the charge transfer current.

For a particle with a surface normal $\hat{\mathbf{n}}$, this is given by matching the flux with the interfacial current density,
\begin{align}
    \mathbf{F}_s \cdot \hat{\mathbf{n}} = -\frac{j}{F} 
    \label{eqn:particlebc}
\end{align}
The average concentration in the particle, $\langle c_s \rangle_p$, only depends on the integrated interfacial charge transfer current density and is given by,
\begin{align}
    \frac{\partial  \langle c_s \rangle_p}{\partial t} = \frac{\int_{A_p} j dA}{F} \frac{1}{V_p} 
    \label{eqn:spbalinteg}
\end{align}
where $A_p$ and $V_p$ denote electrochemically active area and volume of the particle.
The integral in RHS must be evaluated at points on the surface where the concentration is a solution to Equation \ref{eqn:particledyn} and is generally not uniform.

For a fixed particle geometry and interfacial kinetics, the description of $\mathbf{F}_s$ influences the dynamic concentration profile on the electrochemically active area in the RHS of Equation \ref{eqn:spbalinteg}.
Typically assumed to obey Fick's law, $\mathbf{F}_s$ can be more complex depending on the non-equilibrium thermodynamic description of quasi-neutral polaron transport in the electrode material \cite{islam2014lithium}.
This introduces significant complexities requiring microscopic description of the transport in the electrode material having contributions from coupled anisotropic concentration, interfacial energy, strain, and temperature effects \cite{cogswell2012coherency, amin2016characterization}.

Equation \ref{eqn:spbalinteg} is greatly simplified in the limit of fast intra-particle transport (reaction-limited electrode) as the average concentration ($\langle c_s \rangle_p$) collapses to the concentration $c_s$ in the particle with Equation \ref{eqn:spbalinteg} now replaced by,
\begin{align}
    \frac{\partial  c_s}{\partial t} \approx \frac{j}{F} \frac{A_p}{V_p}
    \label{eqn:spbalintegsimple}
\end{align}
where $c_s$ can still depend on the electrode position, $x$, through electrolyte polarization in $j$. 
Seemingly restrictive at first glance, there is growing evidence to suggest that LIB electrodes may be reaction-limited in practice.
For instance, reaction-limited single crystal NMC particles exhibiting size-dependent fictitious phase transformations \cite{park2021fictitious}, agglomerate electrodes with electrolyte-infiltration enhanced transport \cite{trevisanello2021polycrystalline} and sub-micron LFP platelets well described by the Allen-Cahn reaction model, which neglects solid-state diffusion in the active material \cite{zhao2023learning, ombrini2025modeling}.

Importantly, this reaction-limited approximation removes the intra-particle radial coordinate $r$, the pseudo-dimension of the standard pseudo-two-dimensional (P2D) model, as an independent variable. The volume-averaged solid concentration $c_s$ becomes a function of the electrode coordinate $x$ and time $t$ alone, so that the model is reduced from a pseudo-two-dimensional to a genuinely one-dimensional description. This dimensional reduction underlies the closed-form solutions derived in the following sections.

\section{Scaling Analysis} \label{sec:scalingsec}

A critical step in simplifying PET and revealing its mathematical structure is the identification of dimensionless groups that govern the solutions, based on the characteristic scales for length, time, potential, and concentrations. Consider an electrode of thickness $L$, nominal porosity $\epsilon_p$, active material fraction $\epsilon_{am}$, electrolyte concentration $c_{l,\text{ref}}$, potential $\phi_\text{ref}$, maximum solid concentration $c_{s,\text{max}}$, and 
let $t_p$ be the process timescale (e.g., 1 hour for 1 C discharge).
The following dimensionless variables, denoted by a tilde (\textasciitilde), can then be defined based on characteristic scales for length ($\tilde{x} = x/L$), time ($\tilde{t} = t/t_p$), particle concentration ($\tilde{c}_s = c_s/c_{s,\text{max}}$), electrolyte concentration $\tilde{c}_l = c_l/c_{l,\text{ref}}$ and potentials $\tilde{\phi}_l = \phi_l/\phi_{\text{ref}}$, $\tilde{\phi}_s = \phi_s/\phi_{\text{ref}}$. While various choices are possible in computation, the most natural potential scale is the thermal voltage, $\phi_{\text{ref}}=k_BT/e = R_gT/F$, which arises in all models of thermally activated reaction kinetics. Therefore, thermal voltage is chosen as the potential scale henceforth.

With these definitions, we arrive at the following systems of dimensionless equations for a reaction-limited porous electrode,
\begin{align}
    \frac{\partial \tilde{c}_s}{\partial \tilde{t}} &= Da_p \tilde{j} \label{eqn:csbal} \\ 
   \tilde{\tau}_l\frac{\partial \tilde{c}_l}{\partial \tilde{t}} &=  \frac{\partial}{\partial \tilde{x}}\left(\tilde{D}_{\text{eff}}\frac{\partial \tilde{c}_l}{\partial \tilde{x}}\right)-Da \left(\tilde{j} + Da_c^{-1} \frac{\partial \Delta\tilde{\phi}}{\partial \tilde{t}}\right) {} \label{eqn:clbal} \\ 
    \frac{\partial^2 \Delta\tilde{\phi}}{\partial \tilde{x}^2} &= - Da_w\left(\tilde{j} - Da_c^{-1} 
\frac{\partial \Delta\tilde{\phi}}{\partial \tilde{t}}\right) \label{eqn:delphibal}
\end{align}
where $\tilde{j} = j/j_0$ for a reference current per active area $j_0$, $\tilde{D}_{\text{eff}} = {D_{\text{eff}}}/{D_{\text{ref}}}$ for a reference electrolyte diffusivity $D_{\text{ref}} = D(c_{l,\text{ref}})$.
$\Delta\phi = \phi_l - \phi_s$ is defined to combine the two charge balance equations into one equation (Appendix \hyperref[sec:apdxa1]{A.1}). 

Five physically meaningful dimensionless groups appear naturally,
% I SUGGEST: Everywhere (also above): 
% replace F with e, and explicitly write k_BT/e for phi_ref
\begin{align}
Da &= \frac{L^2 j_0 a_p (1-t_+)}{F \epsilon_p D_{\text{ref}} c_{l,\text{ref}}}  \label{eq:Da} \\
Da_p &= \frac{t_p j_0 a_p}{ \epsilon_{am} F c_{s,max}} \\
    Da_w &= \frac{FL^2 j_0 a_p}{R_gT \sigma_{\text{eff}}} \\ 
    Da_c &= \frac{ Fj_0t_p }{R_gT C_{DL}} \\ 
    \tilde{\tau}_l &= \frac{L^2}{D_{\text{ref}}t_p}
\end{align}
where the first four compare various porous-electrode time scales to the characteristic reaction time, set by the exchange current per volume, $j_0 a_p$: (i) $Da$ is a traditional Damköhler number, defined as the ratio of the reaction rate to a characteristic diffusion rate in the electrolyte; (ii) $Da_p$ is the ``process'' Damköhler number, which compares the reaction rate with the rate of capacity utilization (C-rate) set by the applied current; (iii) $Da_w$ is the ``wiring Damköhler number,'' defined as the ratio of the reaction rate to an effective charge transport rate, based on the combined electronic and ionic conductivity, 
% CHECK: I CHANGED a_i to a_p
\begin{align}
{\sigma}_{\text{eff}} = \left(\sigma_s^{-1} + \kappa_l^{-1} + \frac{2R_gT}{FL^2 j_0 a_p}(1-t_+) Da\right)^{-1}, \label{eq:sigma}
\end{align}
which represents ionic, electronic and Faradaic charge transfer processes in series; and (iv) $Da_c$ is the ``capacitive Damköhler number, which compares the reaction rate to the rate of capacitive charging of the electric double layers. Finally, $\tilde{\tau}_l$ is the ratio of the electrolyte diffusion time to the process time. Note that $Da_w$ includes the ionic conductivity $\kappa_l$, which generally depends on the electrolyte concentration, $\tilde{c}_l$. 

% THIS IS THE ONLY NATURAL CHOICE FOR VOLTAGE SCALE, SO USE IT!
%From this point, ${\phi}_{\text{ref}}$ is taken as thermal voltage $k_BT/e$, such that $\tilde{\phi}_{\text{ref}} = 1$, consistent with the scaling of the overpotential in thermally activated reaction models.

Damköhler numbers are widely used in chemical engineering for reactor design \cite{fogler1999elements}, and those defined in Eqs. (\ref{eq:Da})-(\ref{eq:sigma}) could have similar impact on the design of electrochemical systems, well beyond the range of validity of the approximate models derived here. The importance of dimensionless groups is well known in fluid mechanics, where the Reynolds number governs transitions from creeping to turbulent flows, even when mathematical models are too complicated to solve analytically. In electrochemistry, various definitions of the traditional Damköhler number $Da$ have been proposed to scale reaction rates to diffusion rates in fuel cells and flow batteries~\cite{chakrabarti2020modelling,greco2021limited,majji2023modeling,manikandan2025modeling,maggiolo2020solute}, electro-sorption based separations~\cite{clarke2024insights,he2021theory}, lithium-air batteries~\cite{wan2021methods}, electrodeposition~\cite{khoo2019linear,fraggedakis2020tuning}, and voltammetry~\cite{yan2017theory}. Similarly, for  electrocatalysis~\cite{fu2015heterogeneous,wan2021methods,lasia1998hydrogen,lasia2023impedance,lasia2008modeling,halhouli2016sensitivity}, lithium-sulfur batteries~\cite{song2025reducing} and thermoelectrochemical cells~\cite{kim2019diffusion}, the competition of reactions and diffusion in porous electrodes has also been expressed in terms of the equivalent Thiele modulus, $\phi = \sqrt{Da}$. For phase-separating battery materials, Damkohler numbers have been defined that compare intercalation reaction kinetics with solid-state diffusion~\cite{singh2008intercalation,bazant2017thermodynamic,nadkarni2019modeling}, and phase morphologies from diverse experiments and simulations have been collapsed by a universal scaling law~\cite{fraggedakis2020scaling}, which compares $Da$ with a process Damköhler number (ratio of exchange current to applied current) analogous to $Da_p$. A capacitive Damköhler number similar to $Da_c$ has also been defined for reactive porous electrodes in capacitive deionization~\cite{biesheuvel2011diffuse}. Building on these isolated cases, however, our scaling analysis based on four Damköhler numbers ($Da, Da_p, Da_w, Da_c,\tilde{\tau}_l$) covers a much wider range of engineering conditions for porous electrodes.

% his motivates the need for deeper analysis of this limit.
% % On a side note, if intra-particle transport limitations do become important in cases such as large micron-sized particles, model extensions can be made for linear Fickian diffusion by adding corrections using known solutions to the resulting linear partial differential equation for species balance.
% This motivates a closer look at this limit which is the focus of this work.
%Combine all equations and proceed

% \begin{align}
%     \frac{\partial \langle \tilde{c}_s \rangle }{\partial \tilde{t}} + \frac{\partial\delta \tilde{c}_s}{\partial \tilde{t}} = Da_p 
% \end{align}

% \begin{align}
%     \frac{\partial \delta \tilde{c}_s}{\partial \tilde{t}} \approx Da_p \frac{\partial\tilde{j}}{\partial\tilde{c}_s}\delta\tilde{c}_s + Da_p \frac{\partial \tilde{j}}{}
% \end{align}

\section{ Lean Model Equations }

\label{sec:approach}
Equations \ref{eqn:csbal}-\ref{eqn:delphibal} derived in Section \ref{sec:scalingsec} present a system of coupled PDEs that could, in principle, exhibit widely different dynamics depending on the description of the reaction kinetics ($\tilde{j}$) and the electrode material thermodynamics ($\Delta \tilde{\phi}_{eq}$). The functional forms of $\tilde{j}$ and $\Delta \tilde{\phi}_{eq}$ are generally non-linear, which makes general analytical solutions nearly impossible to derive.

In this section, we derive a ``lean model" of simplified dimensionless PET equations, which permits analytical solutions and fast computations, based on two key assumptions:
\begin{enumerate}
    \item \textit{Linear reaction kinetics:}  Charge transfer kinetics is well approximated by a  linearized overpotential dependence, $\tilde{j} \sim -f(\tilde{c}_s,\tilde{c}_l)\tilde{\eta}$, where the exchange current prefactor generally depends nonlinearly on the concentrations and temperature. 
    \item \textit{Fast solid diffusion:} Concentration dependence in $\tilde{j} \propto f(\tilde{c}_s,\tilde{c}_l)$ and $\Delta\phi_{eq}$ are well approximated by $\tilde{c}_s \sim \langle \tilde{c}_s \rangle$ where $\langle.\rangle$ represents a spatial average.
\end{enumerate}
These approximations significantly simplify the PET equations, while retaining remarkable accuracy over a wide range of conditions, as illustrated below.

The first assumption is both practically and theoretically motivated and asserts a nearly constant Faradaic resistance, which depends on reactive species concentrations and temperature. This assumption is supported by experimental measurements of LIB intercalation rates discussed in Section \ref{sec:ctk} and is consistent with the predictions of coupled ion-electron transfer theory over the typical range of activation overpotentials~\cite{zhang2025lithium,bazant2023unified}.

The second assumption considers the deviation of concentration from its spatial average, $\delta \tilde{c}_s = \tilde{c}_s - \langle \tilde{c}_s \rangle $ to be small ($\delta \tilde{c}_s \ll \langle \tilde{c}_s \rangle$).
This is motivated by both mathematical and practical considerations. From a mathematical standpoint, when the electrode behaves like a solid solution such that $\frac{\partial \Delta \phi_{eq}}{\partial \tilde{c}_s} < 0$, the formation of lithiation zones with large boundary concentration gradients is unlikely \cite{bazant2017thermodynamic}.
We show this in Appendix \hyperref[sec:apdxa2]{A.2} by analyzing the leading order solution the governing equations in the limit of negligible electrolyte polarization effects.
From a more practical perspective, experimental maps of internal lithiation profiles in solid-solution electrode materials such as NMC materials consistently show heterogeneities that have significant randomness at particle length scales ($\sim 1 \mu$m) and at thin electrode scales ($\sim 10 \mu$m) \cite{lee2025strain}.
These spatial profiles are not predicted by classical DFN porous electrode models assuming simple models of solid diffusion in spherical particles and liquid electrolyte diffusion in porous electrodes. In contrast, this heterogeneity can be largely captured by reaction-driven population dynamics assuming particles of nearly uniform concentration affected by electro-autocatalysis~\cite{park2021fictitious,bazant2017thermodynamic}.
This work suggests that the behavior of a porous electrode with millions of interacting particles reflects spatially averaged intra-particle concentration profiles, approximately described by $\langle \tilde{c}_s\rangle$ at the porous electrode scale.
It is expected that such an approximation is most realistic for thin ($O(10) \mu$m) electrodes and will need correction terms for transport limitations in progressively thicker electrodes ($> O(100) \mu$m) \cite{singh2015thick, gallagher2016optimizing}. We later compute corrections to deal with this practical limitation.

We are now ready to incorporate these two assumptions, which yields the following system of equations, 
\begin{align}
    \frac{\partial \tilde{c}_s}{\partial \tilde{t}} &= -Da_pf(\tilde{c}_s,\tilde{c}_l)\tilde{\eta} \label{eq:approxsyscs}
\\ 
   \tilde{\tau}_l\frac{\partial \tilde{c}_l}{\partial \tilde{t}} &=  \frac{\partial}{\partial \tilde{x}}\left(\tilde{D}_{\text{eff}}\frac{\partial \tilde{c}_l}{\partial \tilde{x}}\right)-Da \left(-f(\langle \tilde{c}_s \rangle,\tilde{c}_l)\tilde{\eta} + Da_c^{-1} \frac{\partial \Delta\tilde{\phi}}{\partial \tilde{t}}\right) \label{eq:approxsyscl} \\ 
    \frac{\partial^2 \tilde{\eta}}{\partial \tilde{x}^2} &= Da_w\left(f(\langle \tilde{c}_s \rangle,\tilde{c}_l)\tilde{\eta} + Da_c^{-1} \frac{\partial\Delta\tilde{\phi}}{\partial \tilde{t}}\right) \label{eq:approxsysphi}
\end{align}
where we define $\tilde{\eta} = \Delta \tilde{\phi}- \Delta\tilde{\phi}_{eq}$ to reformulate the equations in a more compact form.

For most practical applications, the transport of ions in the electrolyte is much faster than the overall discharge process \cite{marquis2019asymptotic}.
This separation of timescales can be formalized by evaluating $\tilde{\tau}_l\ll 1$, which compares the electrolyte diffusion timescale ($\tau_l = L^2/D_{\text{eff}}$) to the discharge timescale ($t_p$). This justifies neglecting the transient accumulation term for ions in the electrolyte, $\partial c_l/\partial t$, which we assume for simplicity hereafter.

\section{ Analytical Solutions of the Lean Model }

\subsection{Constant Current } 
\label{sec:approxanalytical}

We will first derive approximate solutions for galvanostatic (dis)charge.
Consider a galvanostatic discharge with a specified C-rate. Then, the process timescale (in seconds) is naturally given by, $t_p \sim \frac{3600}{\text{C-rate}}$.
For typical battery discharge, $t_p$ is on the order of $\mathcal{O}$(10 min) \cite{bard2022electrochemical, li2023characterizing}.
This is significantly longer than double layer charging and electrolyte transport timescales so that $Da_c^{-1}, \tilde{\tau}_l \ll 1$.
Therefore, to a first approximation, capacitive and accumulation terms in Equations \ref{eq:approxsyscs}-\ref{eq:approxsysphi} can be dropped. The final set of approximated governing equations is
\begin{align}
  \frac{\partial \tilde{c}_s}{\partial \tilde{t}} &= -Da_pf(\tilde{c}_s,\tilde{c}_l)\tilde{\eta} \\
    \frac{\partial^2 \tilde{c}_l}{\partial \tilde{x}^2} &\approx  -Daf(\langle \tilde{c}_s \rangle,\tilde{c}_l)\tilde{\eta}  \\ 
    \frac{\partial^2 \tilde{\eta}}{\partial \tilde{x}^2} &\approx Da_wf(\langle \tilde{c}_s \rangle,\tilde{c}_l)\tilde{\eta} \label{eq:galvanoxi}
\end{align}
where for small polarization effects, $\tilde{D}_{\text{eff}}$ is approximated as $\tilde{D}_{\text{eff}} \approx 1$. We now analytically compute the solution for $\tilde{\eta}$. Analytical solutions for other related variables like $\tilde{\phi}_s$ are found by recasting their corresponding governing equation in terms of $\tilde{\eta}$, a technique employed in prior analytical impedance studies \cite{devan2004analytical, sikha2008analytical}.

First consider the limit of negligible polarization i.e., $Da \ll Da_p$ for which $\tilde{c}_l \approx \tilde{c}_{l,0}$.
Equation \ref{eq:galvanoxi} becomes a second order linear ODE with a Neumann boundary condition for current density at the current collector, $\frac{\partial \tilde{\eta}}{\partial \tilde{x}}|_{\tilde{x}=1} = -\frac{Da_{w,\sigma}}{Da_p}$.
The solution can be expressed in terms of hyperbolic cosine functions as,

\begin{align}
    \tilde{\eta} = \frac{(\langle \tilde{\eta} \rangle \Lambda^2 + Da_{w,\sigma}/Da_p)\cosh(\Lambda (\tilde{x}-1)) -  Da_{w,\sigma}/Da_p\cosh(\Lambda \tilde{x})}{\Lambda\sinh \Lambda } 
    \label{eq:xisolgalvano}
\end{align}
where $\Lambda = \sqrt{Da_w f(\langle \tilde{c}_s\rangle)}$ captures the combined impact of filling fraction dependent kinetics and wiring on the impedance. $\Lambda$ admits a natural interpretation as an eigenvalue of the lean model. Writing the second-order equation $\tilde{\eta}'' = Da_w f\,\tilde{\eta}$ as a first-order system in $(\tilde{\eta},\tilde{\eta}')$, the associated matrix has eigenvalues $\pm\Lambda$, and these characteristic roots dictate the spatial structure of the solution. When $\Lambda$ is real and nonzero, the eigenvalues are distinct and produce the hyperbolic $\cosh/\sinh$ profiles above, with $1/\Lambda$ measuring the characteristic depth of electrode utilization. In the limit $\Lambda \to 0$ the eigenvalues coalesce to zero and the overpotential becomes uniform in a uniformly accessible electrode.

$Da_{w,\sigma}=FL^2 j_0 a_p / (R_g T \sigma_s)$ is the solid phase wiring Damköhler number and the process timescale, $t_p$ in $Da_p$ is $t_p=3600/\text{C-rate}$ seconds.
The spatially averaged overpotential, $\langle \tilde{\eta} \rangle$ is set by the current constraint,

\begin{align}
    \Big\langle \frac{\partial \tilde{c}_s}{\partial \tilde{t}} \Big\rangle  = -Da_p f(\langle \tilde{c}_s\rangle) \langle \tilde{\eta} \rangle \\ 
    \implies \langle \tilde{\eta}\rangle = -\frac{1}{Da_pf(\langle \tilde{c}_s\rangle)}
\end{align}

The voltage across a half cell is given by $V_{cell} = {\phi}_s(\tilde{x}=1) - {\phi}_{ref}(\tilde{x}=0)$.

This can be found by recasting the governing equation for $\phi_s$ (Appendix~\hyperref[sec:apdxa1]{A.1}) in terms of $\eta$, 
\begin{align}
    \frac{\partial^2 \tilde{\phi}_s}{\partial \tilde{x}^2} = \frac{Da_{w,\sigma}}{Da_w} \frac{\partial^2 \tilde{\eta}}{\partial \tilde{x}^2}
\end{align}

Using consistent boundary conditions of current density at current collector, $\frac{\partial \tilde{\phi}_s}{\partial \tilde{x}}|_{\tilde{x}=1} = -\frac{Da_{w,\sigma}}{Da_p}$ and no electronic current at separator, $\frac{\partial\tilde{\phi}_s}{\partial \tilde{x}}|_{\tilde{x}=0} = 0$ results in an expression of $\phi_s$ in terms of $\eta$ as,
\begin{align}
\label{eq:phisintermsofeta}
    \tilde{\phi}_s = \left(\frac{Da_{w,\sigma}}{Da_w}(\tilde{\eta}-\tilde{\eta}'|_{\tilde{x}=0}\tilde{x}) + \left(1-\frac{Da_{w,\sigma}}{Da_w}\right)(\tilde{\eta}|_{\tilde{x}=0} + \Delta \tilde{\phi}_{eq}(\langle \tilde{c}_s \rangle))\right)
\end{align}

The voltage curves are obtained from simply plugging $\tilde{x}=1$ in Equation \ref{eq:xisolgalvano},
\begin{align}
     \tilde{\phi}_{s,app} = \left(\frac{Da_{w,\sigma}}{Da_w}(\tilde{\eta}|_{\tilde{x}=1}-\tilde{\eta}'|_{\tilde{x}=0}) + \left(1-\frac{Da_{w,\sigma}}{Da_w}\right)(\tilde{\eta}|_{\tilde{x}=0} + \Delta \tilde{\phi}_{eq}(\langle \tilde{c}_s \rangle))\right)
     \label{eq:xizerogalvano}
\end{align}
Equation \ref{eq:xizerogalvano} will be used for galvanostatic (dis)charge voltage ($\Delta {\phi}_{app}$) predictions in Section \ref{sec:results}.

\subsection{ Voltage Pulse } 

A variety of experimental protocols apply current pulses (e.g. HPPC) \cite{li2023characterizing} or voltage pulses (e.g., CV hold, chronoamperometry) \cite{bard2022electrochemical}.
The analysis for a constant current pulse largely follows the galvanostatic protocol, so we focus on the current response to voltage pulse.

The approximate governing equations for a voltage pulse are given by,
\begin{align}
    \frac{\partial \tilde{c}_s}{\partial \tilde{t}} \approx -Da_p f \tilde{\eta} \label{eqn:cbalchrono}\\ 
    \frac{\partial^2 \tilde{\eta}}{\partial \tilde{x}^2} \approx Da_w\left(f\tilde{\eta} +Da_c^{-1} \frac{\partial \Delta \tilde{\phi}}{\partial \tilde{t}} \right)
    \label{eqn:etabal}
\end{align}
where we have assumed that the pulses are not severe enough to cause significant electrolyte polarization ($\tilde{c}_l \sim 1$) and $f = f(\langle \tilde{c}_s \rangle,1)$.
Taking the typical limit of long ($\mathcal{O}$(10 min)) chronoamperometry timescales, for which $Da_c^{-1} \cdot Da_w \ll 1$, the pseudo-steady solution to Equation \ref{eqn:etabal} is given by,
\begin{align}
    \tilde{\eta} = \tilde{\eta}_0 (\tilde{t})\frac{\cosh(\Lambda(\tilde{x}-1))}{\cosh \Lambda}
\end{align}
where $\tilde{\eta}_0(\tilde{t}) = \Delta \tilde{\phi}_{app} - \Delta \tilde{\phi}_{eq}(\langle \tilde{c}_s \rangle)$ for a fixed applied potential $\Delta \phi_{app}$ and $\Lambda = \sqrt{Da_w f(\langle \tilde{c}_s\rangle)}$.
Plugging this in Equation \ref{eqn:cbalchrono}, and taking an average yields the scaled current, $\tilde{I} = \frac{\partial \langle \tilde{c}_s \rangle}{\partial \tilde{t}}$ as,
\begin{align}
    \frac{\partial \langle \tilde{c}_s \rangle} {\partial \tilde{t}} \approx -Da_p f(\langle \tilde{c}_s \rangle) \cdot (\Delta \tilde{\phi}_{app} - \Delta \tilde{\phi}_{eq}(\langle \tilde{c}_s \rangle))\frac{\tanh \Lambda}{\Lambda}
\end{align}
The above equation can be compactly written as,
\begin{align}
    \frac{\partial \langle 
    \tilde{c}_s \rangle}{\partial \tilde{t}} = \mathcal{X}(\langle 
    \tilde{c}_s \rangle)
        \label{eqn:chronoampeqraw}
\end{align}
for $\mathcal{X}(\langle  \tilde{c}_s \rangle) = -Da_p f(\langle \tilde{c}_s \rangle) \cdot (\Delta \tilde{\phi}_{app} - \Delta \tilde{\phi}_{eq}(\langle \tilde{c}_s \rangle))\frac{\tanh \Lambda}{\Lambda}$.

While complete analytical treatment of Equation \ref{eqn:chronoampeqraw} depends on the functional form of $\mathcal{X}(.)$, in most experiments the concentration change in a pulsing measurement is small.
Therefore, a linear approximation about equilibrium concentration, $\mathcal{X}(\langle \tilde{c}_s \rangle)  \approx \frac{\partial \mathcal{X}}{\partial \tilde{c}_s}\Big|_{\tilde{c}_{s,eq}} (\langle \tilde{c}_s\rangle - \tilde{c}_{s,eq})$, is sufficient.
This yields an exponentially decaying current which is given by,
\begin{align}
    \tilde{I} = \frac{\partial \langle \tilde{c}_s \rangle}{\partial \tilde{t}} \approx \mathcal{X}'(\tilde{c}_{s,eq})(\tilde{c}_{s,0} - \tilde{c}_{s,eq}) \exp(\mathcal{X} '(\tilde{c}_{s,eq}) \tilde{t})
    \label{eqn:chronoampeq}
\end{align}
where the initial concentration is $\tilde{c}_{s,0}$.

\subsection{ Impedance } 

Finally, we derive the approximate electrochemical impedance spectrum (EIS) of a porous electrode. Since impedance analysis assumes linear response to sinusoidal forcing, no  additional approximations are necessary. The impedance is defined as the ratio of complex voltage and current amplitudes,  $Z(\omega)= \frac{\phi^*}{I^*}$, for a sinusoidal voltage perturbation, $\delta \phi = \phi^*e^{j\omega t}$, in response to a sinusoidal current perturbation, $\delta I=I^*e^{j \omega t}$.
For small current perturbation amplitudes $I^*$, electrolyte polarization effects may be neglected, $\tilde{c}_l \sim 1$.

Accordingly, the key perturbed equations in this case are,
\begin{align}
    j\tilde{\omega} \tilde{c}_s^* &\approx -Da_p f \tilde{\eta}^* \label{eqn:cbalstar}\\ 
    \frac{\partial^2 \Delta \tilde{\phi}^*}{\partial \tilde{x}^2} &\approx Da_w(f \tilde{\eta}^* + Da_c^{-1} j\tilde{\omega} \Delta \tilde{\phi}^*)  \label{eqn:xibalstar}
\end{align}
where complex amplitudes for perturbed variables are $\tilde{c}_s^*$, $\tilde{\eta}^*$, $\Delta \phi_{eq}^*$ and $\Delta \phi^*$ and $\tilde{\omega} = t_p \omega$ is a dimensionless frequency. The choice of $t_p$ here is arbitrary since the process timescale is already set by the perturbation frequency $1/\omega$ we can set $t_p=1$ second and drop the tilde ( $\tilde{}$ ) on frequency for derivations beyond this point.
$f = f(\tilde{c}_s,\tilde{c}_l)\big|_{ss}$ is the value of the concentration dependent pre-factor at steady state.

Using the definition of $\tilde{\eta} = \Delta \tilde{\phi}- \Delta \tilde{\phi}_{eq}$, the complex amplitudes can be expressed as, 
\begin{align}
    \tilde{\eta}^* = \Delta \tilde{\phi}^* -  \frac{\partial \Delta\tilde{\phi}_{eq}}{\partial\tilde{c}_s}\Big|_{\tilde{c}_{s,ss}}\tilde{c}_s^* \label{eqn:xistar}
\end{align}
Rearranging Equation \ref{eqn:xistar} and differentiating in $\tilde{x}$ twice yields,
\begin{align}
     \frac{\partial^2 \tilde{\eta}^*}{\partial \tilde{x}^2} &= \frac{Da_w}{1-\frac{\partial \Delta \tilde{\phi}_{eq}}{\partial \tilde{c}_s}\frac{Da_p f}{j\omega}}\left(f \tilde{\eta}^* + Da_c^{-1} j\omega \left(\tilde{\eta}^* + \frac{\partial \Delta\tilde{\phi}_{eq}}{\partial\tilde{c}_s}\Big|_{\tilde{c}_{s,ss}} \tilde{c}_s^*\right)\right)  \label{eqn:xibalstar}
\end{align}
Equations \ref{eqn:cbalstar} and \ref{eqn:xibalstar} are second-order linear PDEs that can be solved analytically to obtain
\begin{align}
    \tilde{\eta}^* =\frac{\Xi_\kappa\cosh(\Omega (\tilde{x}-1)) +  \Xi_\sigma
    \cosh(\Omega \tilde{x})}{\Omega\sinh \Omega}   \label{eqn:xisolstar}
\end{align}
with,
\begin{align}
    \Omega = 
\sqrt{\frac{Da_w}{1-\frac{\partial \Delta \tilde{\phi}_{eq}}{\partial \tilde{c}_s}\frac{Da_p f}{j\omega}}\left(f + Da_c^{-1}j\omega -\frac{Da_pf}{Da_c}\frac{\partial \Delta\tilde{\phi}_{eq}}{\partial\tilde{c}_s}\Big|_{\tilde{c}_{s,ss}}\right)  }
\end{align}
where the current matching boundary conditions 
\begin{align}
    \frac{\partial\tilde{\eta}^*}{\partial \tilde{x}}\Big|_{\tilde{x}=1} = \Xi_\sigma = -\tilde{I}^*\frac{Da_{w,\sigma}}{Da_p}\frac{1}{1-\frac{\partial \Delta \tilde{\phi}_{eq}}{\partial \tilde{c}_s}\frac{Da_p f}{j\omega}}
\end{align} and 
\begin{align}
    -\frac{\partial\tilde{\eta}^*}{\partial \tilde{x}}\Big|_{\tilde{x}=0} = \Xi_\kappa = -\tilde{I}^*\frac{Da_{w,\kappa}}{Da_p}\frac{1}{1-\frac{\partial \Delta \tilde{\phi}_{eq}}{\partial \tilde{c}_s}\frac{Da_p f}{j\omega}}
\end{align}

% The current density, ${I}^*$ at $\tilde{x}=0$ may be approximated as,
% \begin{align}
%     I^*= \frac{L\epsilon_{am}Fc_{s,max}}{t_p}\frac{Da_p}{Da_w}\frac{\partial \Delta \tilde{\phi}^*}{\partial \tilde{x}}\Big|_{\tilde{x}=0} \label{eqn:istar}
% \end{align}
The impedance of a half cell is,
\begin{align}
    Z = \frac{\phi_s^*\big|_{\tilde{x}=1} - {\phi}_{Li}^*}{I^*}\label{eqn:zdef}
\end{align}
where $\phi_{Li}^* = \phi_l^*|_{\tilde{x}=0} = 0$ for infinitely fast charge transfer kinetics at Li anode.
Comparing the governing equation for $\phi_s$ (Appendix~\hyperref[sec:apdxa1]{A.1}) to that for $\eta^*$ in Equation \ref{eqn:xibalstar} leads to the following relationship,
\begin{align}
    \frac{\partial^2 \tilde{\phi}_s^*}{\partial \tilde{x}^2} = \left(1-\frac{\partial \Delta \tilde{\phi}_{eq}}{\partial \tilde{c}_s}\frac{Da_p f}{j\omega}\right)\frac{Da_{w,\sigma}}{Da_w} \frac{\partial^2 \tilde{\eta}^*}{\partial \tilde{x}^2}
\end{align}
The solution is given by,
\begin{align}
    \tilde{\phi}_s^* = \left(1-\frac{\partial \Delta \tilde{\phi}_{eq}}{\partial \tilde{c}_s}\frac{Da_p f}{j\omega}\right)\left(\frac{Da_{w,\sigma}}{Da_w}\left(\tilde{\eta}^* + \Xi_\kappa \tilde{x}\right) + \left(1-\frac{Da_{w,\sigma}}{Da_w}\right) \tilde{\eta}^*\big|_{\tilde{x}=0}\right)
\end{align}
where boundary conditions of zero electronic current at separator, $\frac{\partial \tilde{\phi}_s^*}{\partial \tilde{x}}\big|_{\tilde{x}=0}=0$ and $\phi_l^*\big|_{\tilde{x}=0} =0$ were used.

Substituting Equations \ref{eqn:xistar} in Equation \ref{eqn:zdef},
\begin{align}
     Z = \frac{t_pR_gT}{L\epsilon_{am}F^2c_{s,max}}\left(1-\frac{\partial \Delta \tilde{\phi}_{eq}}{\partial \tilde{c}_s}\frac{Da_p f}{j\omega}\right)\frac{\frac{Da_{w,\sigma}}{Da_w}(\tilde{\eta}^*|_{\tilde{x}=1}+\Xi_\kappa) + \left(1-\frac{Da_{w,\sigma}}{Da_w}\right)\tilde{\eta}^*|_{\tilde{x}=0}}{\tilde{I}^*} 
     \label{eqn:impedanceq}
\end{align}
Substituting the analytical solution for $\tilde{\eta}^*$ in \ref{eqn:xisolstar} yields the desired expression for area-normalized impedance,
\begin{align}
    Z = Z_{\text{ref}} \frac{\Xi_\kappa \Xi_\sigma\left(1 + \frac{2}{\Omega \sinh \Omega}\right) +  (\Xi_\sigma^2+\Xi_\kappa^2) \frac{\cosh \Omega}{\Omega \sinh \Omega}}{(\Xi_\kappa + \Xi_\sigma)^2}  
\end{align}
where $Z_{\text{ref}}$ is the identified scaling for $Z$, combining reaction and electrode wiring resistances, and is given by
\begin{align}
    Z_{\text{ref}} = \frac{Da_w}{Da_p}\cdot \frac{R_gT}{F}\cdot\frac{t_p}{FL\epsilon_{am}c_{s,max}}
\end{align}

\section{ Accuracy of the analytical approximations } \label{sec:results}

In this section, we demonstrate the predictive capabilities of our analytical approximations. We begin by comparing our model predictions against the building blocks of (dis)charging protocols of Li-ion batteries.
These are typically galvanostatic or constant current (CC) and, potentiostatic or constant voltage (CV) steps \cite{bard2022electrochemical}. Following this, we compare the analytical solution in the frequency domain. This is done by comparing Nyquist curves obtained from the analytical formula against simulated frequency domain data for a NMC532 half-cell. Finally, we also visualize limits for which the approximations diverge from simulations. An open access codebase for simulating all the results can be found on Github \cite{codebase}.

To validate the example analytical solutions derived in Section~\ref{sec:approxanalytical}, we compare model predictions against full numerical simulations of Li-ion batteries using two different software packages, MPET~\cite{smith2017multiphase} and PyBaMM~\cite{sulzer2021python}.
Both simulators numerically solve a pseudo-two-dimensional (P2D) model with finite volume spatial discretization and implicit time-stepping \cite{fuller1994simulation, doyle1993modeling}.
The benchmark is built from default MPET electrode parameters for NMC532 to represent a practical 100 $\mu$m thick half-cell with 500 nm spherical single crystal particles and CIET kinetics at the electrode-electrolyte interface.
The complete set of baseline simulation parameters are in Appendix~\hyperref[sec:apdxa6]{A.6}.
Note that although the chosen reaction model is CIET, any model that can be approximated as linear in overpotential during operation would compare well against analytical solutions (e.g., Butler-Volmer (BV)+Film resistance model). The comparisons are made on three canonical electrochemical protocols: galvanostatic discharge (constant current), chronoamperometry (constant voltage), and electrochemical impedance spectroscopy (frequency domain).

\begin{figure}[H]
    \centering
    \begin{subfigure}[t]{0.48\textwidth}
        \centering
        \includegraphics[width=\textwidth]{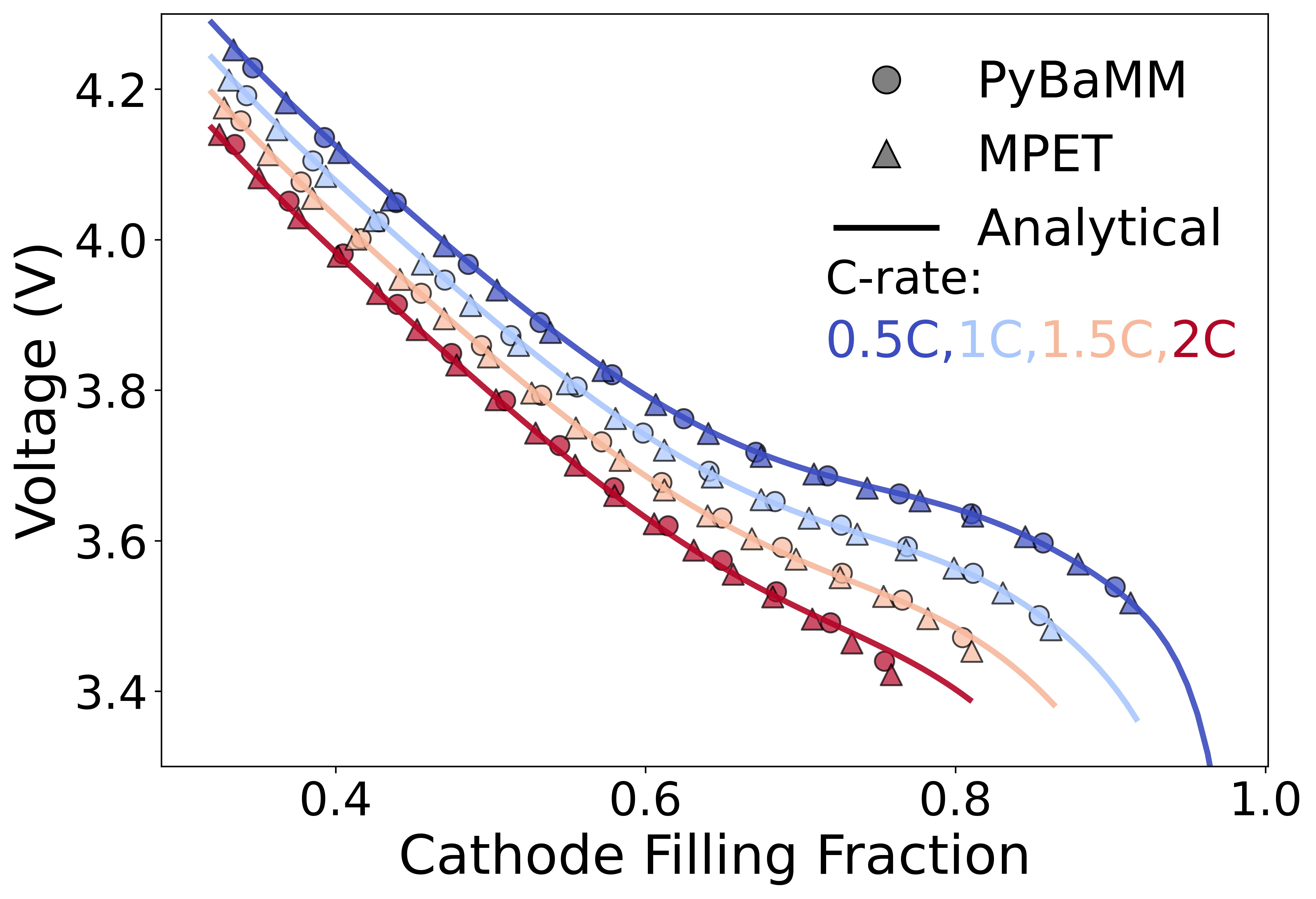}
        % \caption{}
        \label{fig:dischcomp}
    \end{subfigure}
    \hfill  
    \begin{subfigure}[t]{0.48\textwidth}
        \centering
        \includegraphics[width=\textwidth]{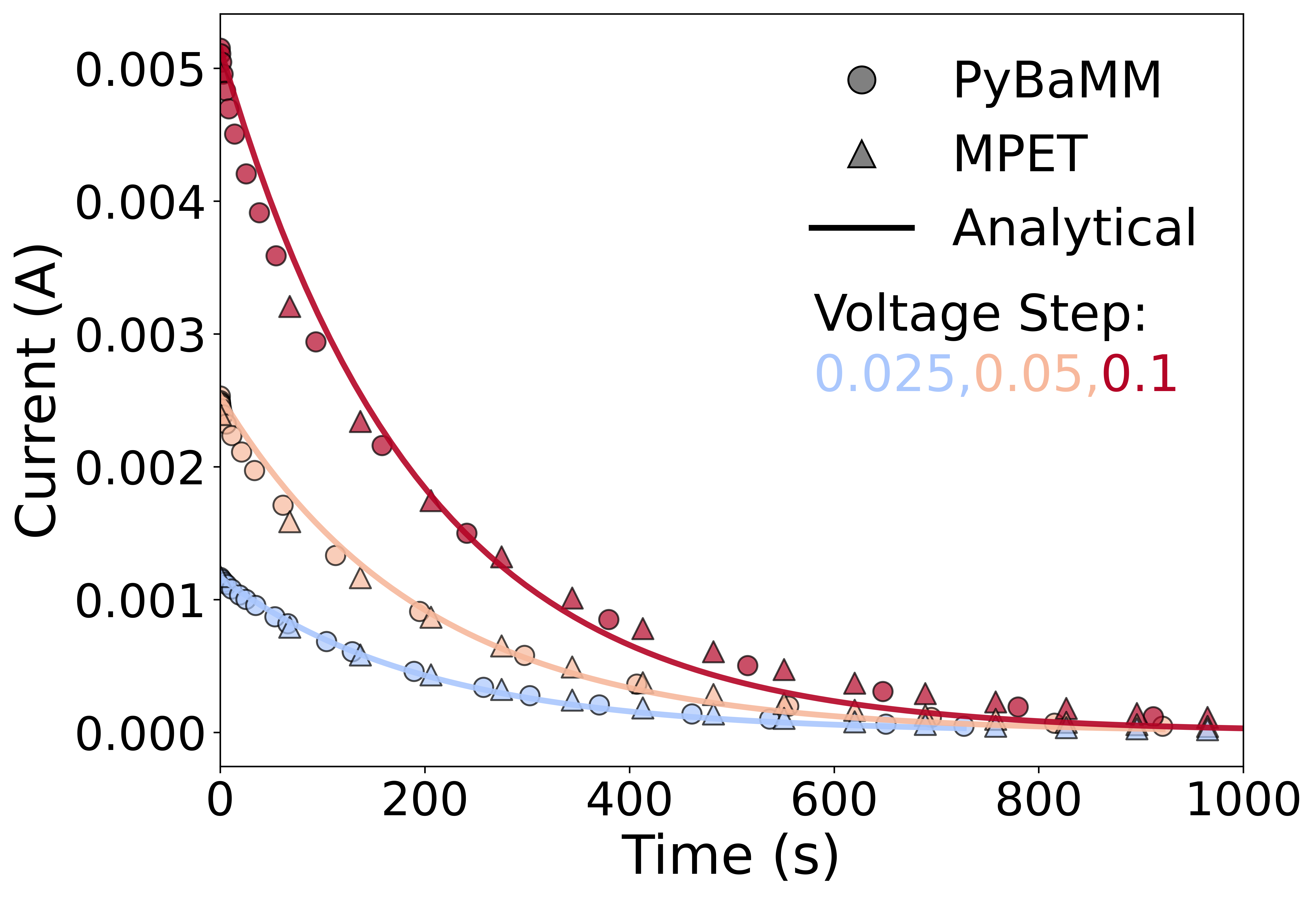}
        % \caption{}
        \label{fig:chronocomp}
    \end{subfigure}
    \vspace{-0.2cm}
     
\caption{Comparison of our analytical approximations with the results of full simulations of the same NMC532 porous electrode using two different software packages (PyBaMM~\cite{sulzer2021python} and MPET~\cite{smith2017multiphase}) for both galvanostatic discharge (left) and chronoamperometry (right) on the same model half-cell.}
    \label{fig:sidebyside}
\end{figure}

Figure~\ref{fig:sidebyside} (left panel) compares the voltage response predicted by the analytical model (Equation~\ref{eq:xizerogalvano}) against numerical simulations for discharge at 0.5 C, 1 C, and 2 C rates.
The analytical approximation accurately captures the voltage across the entire state-of-charge (SOC) range with RMSE of 76.7~mV. At moderate C-rates ($\leq 1$C), the agreement is excellent, with deviations remaining below $\sim$60~mV throughout discharge. At higher rates, slight discrepancies emerge near the end of discharge, where electrolyte transport limitations, neglected beyond leading order in the analytical model, begin to influence cell polarization. This is consistent with the scaling condition derived in Appendix~\hyperref[sec:apdxa2]{A.2}, beyond which the pseudo-steady electrolyte approximation breaks down.

For the same half-cell, an initial galvanostatic discharge to 50\% SOC, a range of discharge voltage pulses of magnitude $\Delta V = \{25,50,100\}$~mV were applied.
Figure~\ref{fig:sidebyside} (right panel) shows the current decay as a function of time. The analytical prediction (Equation \ref{eqn:chronoampeq}) exhibits exponential relaxation.
The predictions show good agreement, with RMSE values of 0.03, 0.06, and 0.13 mA.
The RMSE increases for larger pulses as the linear approximation breaks down.

\begin{figure}[H]
    \centering
    \begin{subfigure}[t]{0.48\textwidth}
        \centering
        \includegraphics[width=\textwidth]{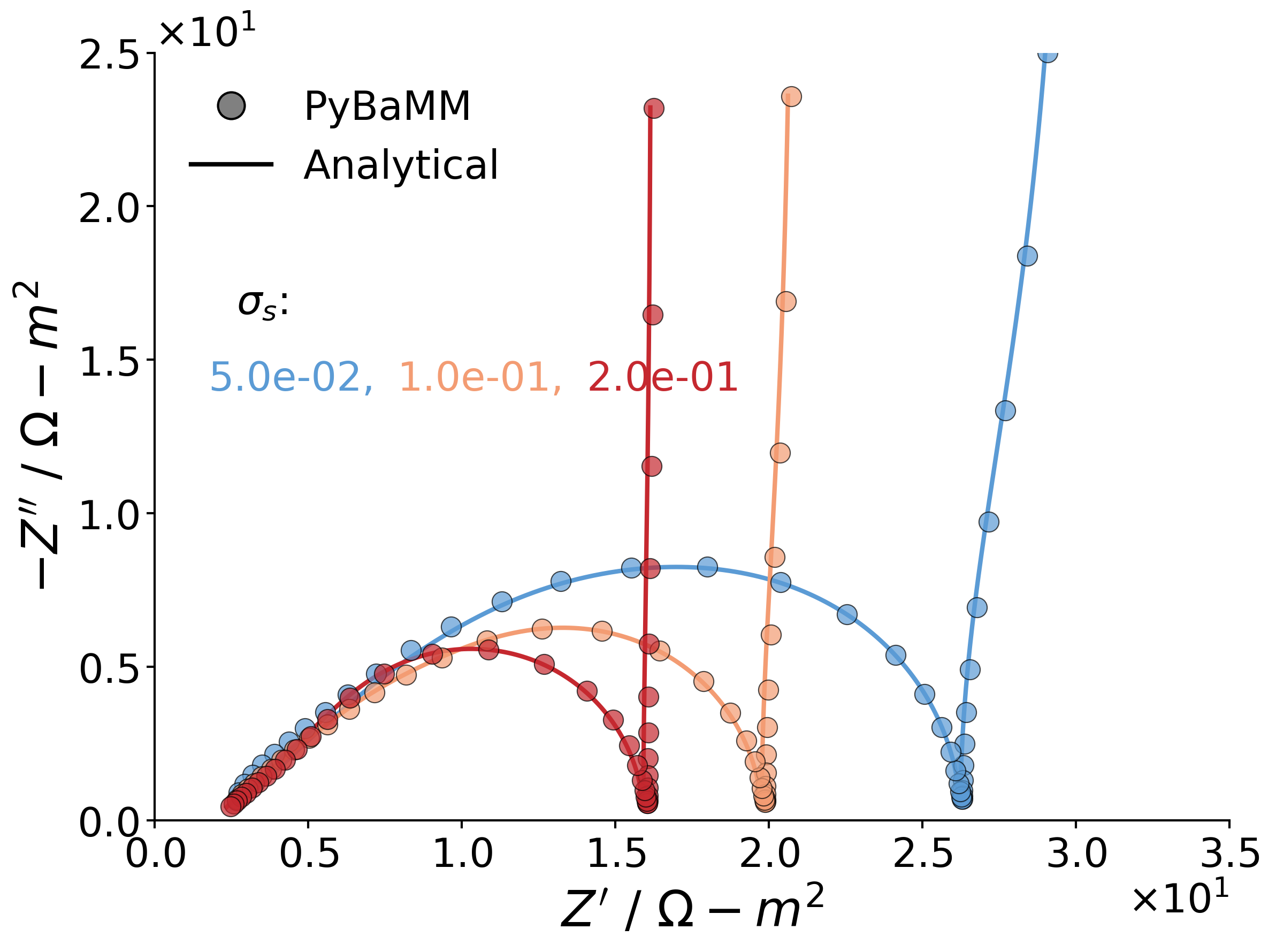}
        % \caption{}
        \label{fig:eisdaw}
    \end{subfigure}
    \hfill
    \begin{subfigure}[t]{0.48\textwidth}
        \centering
        \includegraphics[width=\textwidth]{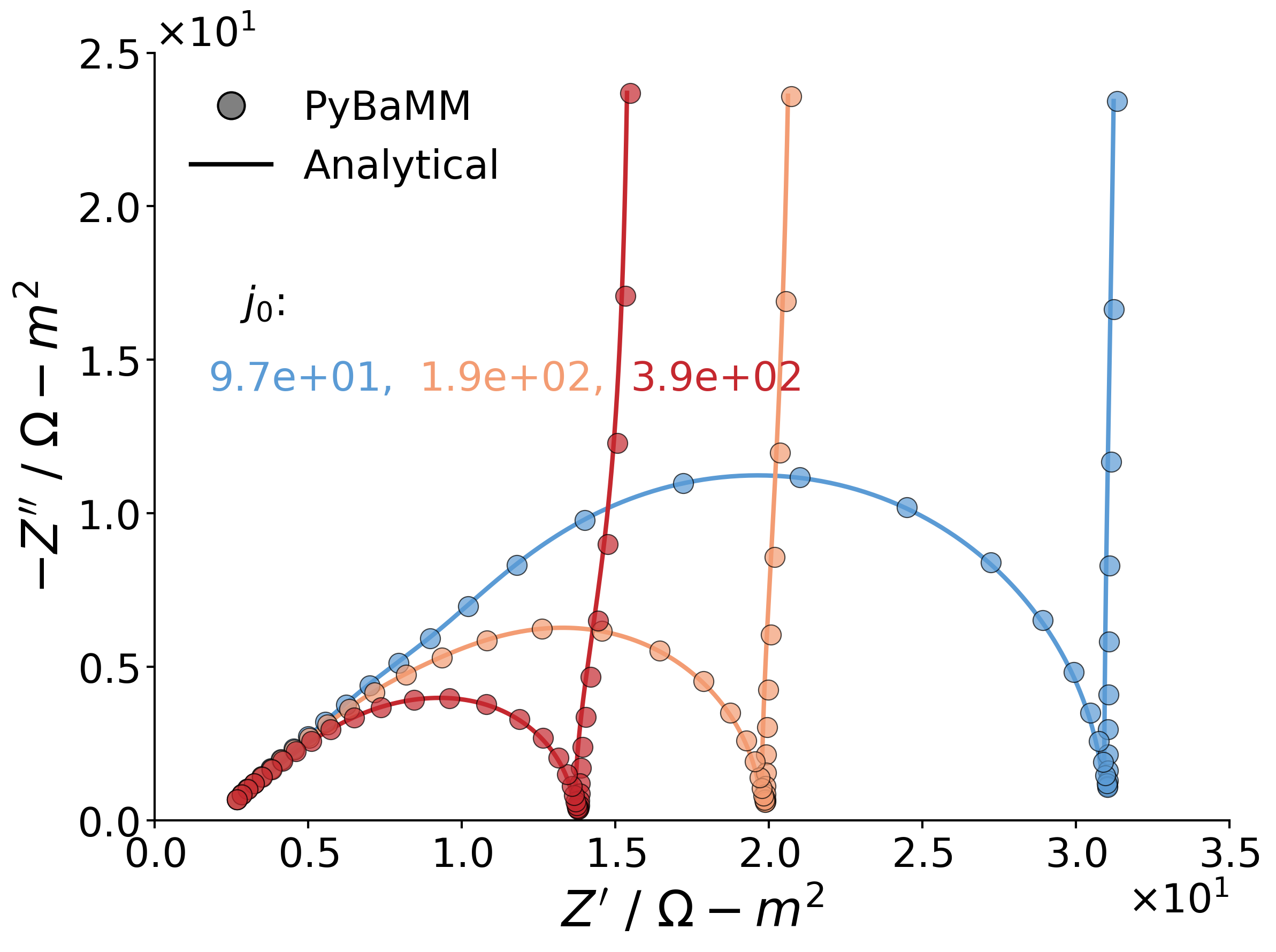}
   
     % \caption{}
        \label{fig:eisdap}
    \end{subfigure}
    \vspace{-0.2cm}
    \caption{Comparison against simulated frequency domain EIS response for changing solid conductivity ($\sigma_s$) (left, RMSE = 0.10 $\Omega$-$m^2$) and exchange current pre-factor ($j_0$) (right, RMSE = 0.13 $\Omega$-$m^2$).
Baseline parameters are in yellow. 
}    \label{fig:sbsEIS}
\end{figure}

Figure~\ref{fig:sbsEIS} presents Nyquist plots comparing the analytical impedance formula (Equation~\ref{eqn:impedanceq}) with simulated frequency-domain data over $10^{-3} \leq \omega \leq 10^3$~Hz.
The analytical model reproduces the characteristic features: a high-frequency intercept corresponding to the ohmic resistance, an arc feature reflecting interfacial charge-transfer kinetics, and a low-frequency tail associated with ionic transport in the electrolyte \cite{orazem2008electrochemical}.
To explore the parametric sensitivity, Figure~\ref{fig:sbsEIS} shows impedance spectra for varying wiring Damköhler number $Da_w$ (left) by changing the solid conductivity $\sigma_s$ and process Damköhler number $Da_p$ by changing the exchange current pre-factor, $j_0$ (right).
Reducing the solid phase conductivity enlarges the semicircle diameter, consistent with higher interfacial resistance.
On the other hand, reducing the exchange current density increases the semicircle diameter and also shifts the characteristic frequency at which the semicircle appears \cite{lasia2023impedance}.
The analytical model correctly captures these trends.

Figures \ref{fig:sidebyside} and \ref{fig:sbsEIS} establish good agreement between our approximations and the simulations for the baseline parameters (Table 1 in Appendix ~\hyperref[sec:apdxa6]{A.6}) and mild protocols. Yet, it is expected that the approximations made in Section \ref{sec:approach} would break under more extreme conditions where concentration polarization and overpotentials are large.
To test the limitations of the approximation, we computed the root-mean-squared error (RMSE) between analytical and simulated voltage curves for galvanostatic discharge. 

\begin{figure}[H]
    \centering
\includegraphics[width=\textwidth]{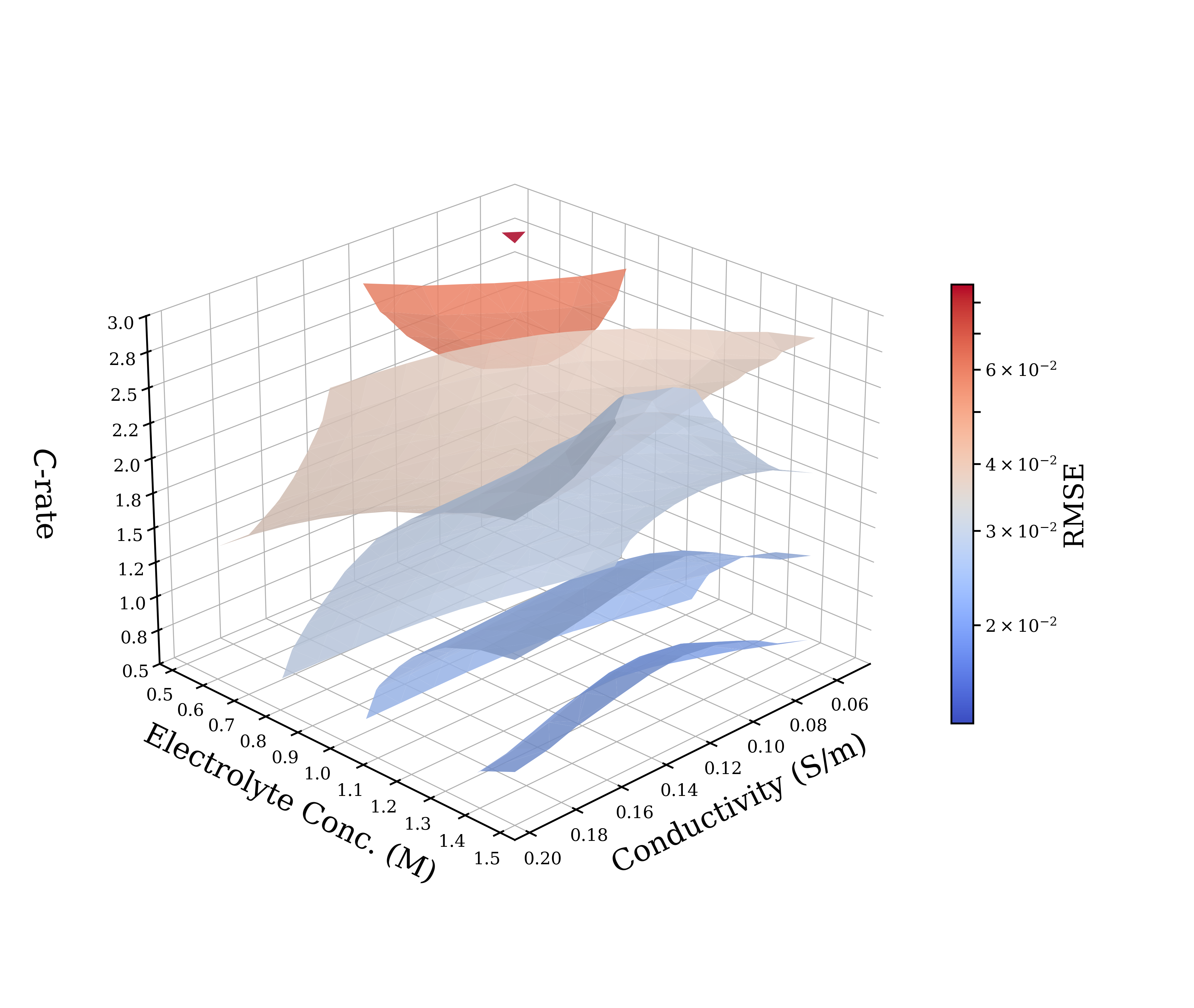}
    \vspace{-0.2cm}
    \caption{Root mean squared error, RMSE (in V) of analytical approximation versus simulated galvanostatic response across C-rates, conductivity ($\sigma_s$) that impacts wiring, and electrolyte concentrations ($c_0$) that impacts polarization.}
    \label{fig:mpetrmse}
\end{figure}

Figure~\ref{fig:mpetrmse} shows RMSE as a function of C-rate, electronic conductivity $\sigma_s$, and initial electrolyte concentration $c_{l,0}$.
The error remains below 0.1 V for most practical operating conditions, increasing primarily in regimes where: (i) high C-rates induce significant electrolyte polarization ($Da \gg Da_p$), violating the pseudosteady assumption;
or (ii) low electronic conductivity ($Da_w \gg Da_p$) causes substantial ohmic losses through the electrode thickness, invalidating the spatially uniform overpotential approximation.

\section{Design Implications} \label{sec:design}

Beyond providing fast, interpretable predictions for a single cell, the dimensionless groups identified in Section~\ref{sec:scalingsec} can be used to define a scaled \emph{design space} for porous electrodes. Because $Da$, $Da_p$, $Da_w$, and $Da_c$ compare the characteristic reaction rate to the rates of electrolyte diffusion, capacity utilization, charge ``wiring,'' and double-layer charging, respectively, the location of a device in this space encodes its dominant physics.

In this section we use the dimensionless groups in two complementary ways: (i) to map representative electrochemical systems onto a design space, and (ii) to derive dimensionless energy and power densities that yield Ragone plots.

\subsection{Design Envelope in Damk\"ohler Space}

\begin{figure}[H]
    \centering
    \includegraphics[width=\textwidth]{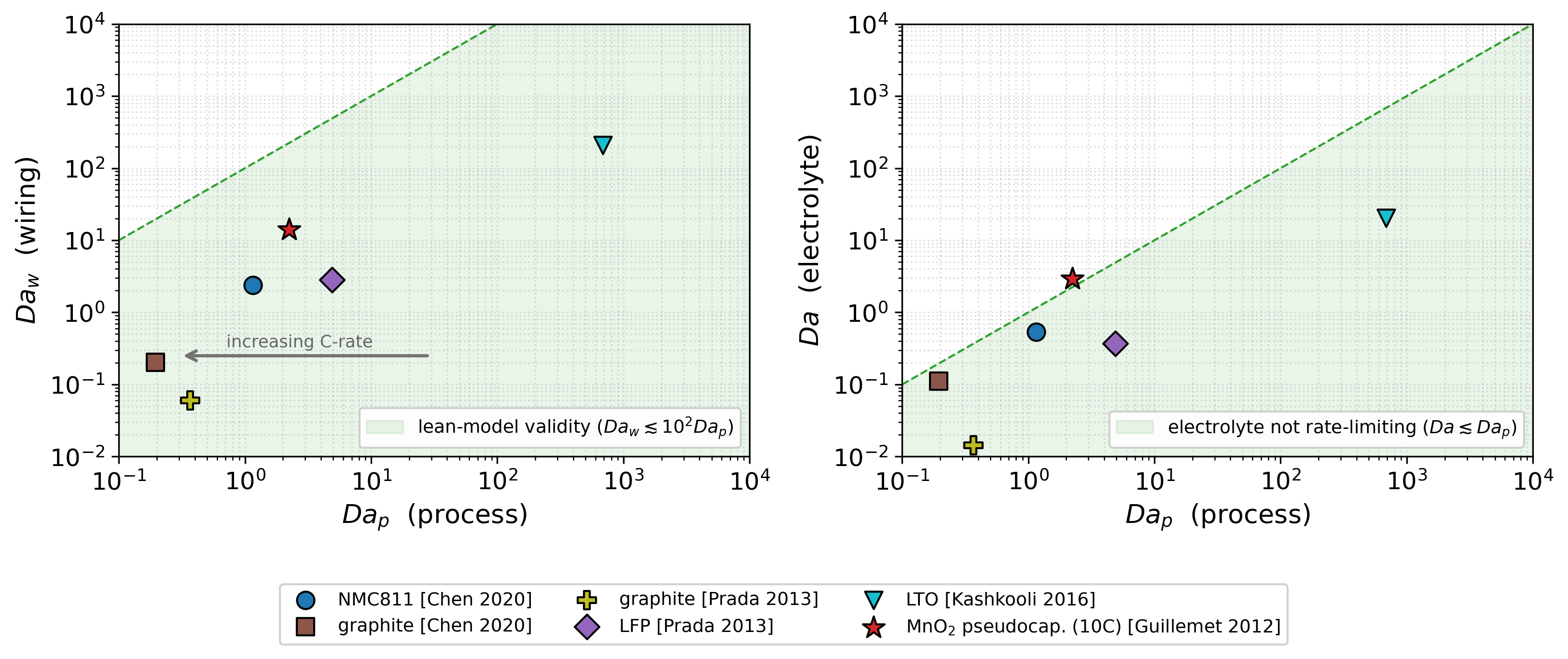}
    \vspace{-0.2cm}
    \caption{Electrochemical electrodes in Damk\"ohler space. \emph{Left:} the $(Da_p, Da_w)$ plane with the lean-model validity region $Da_w \lesssim 10^2 Da_p$ shaded; markers locate three Li-ion positive electrodes- NMC811~\cite{chen2020development}, LFP~\cite{prada2013simplified}, and LTO~\cite{kashkooli2016nano}- and two graphite negative electrodes (from the LG M50 cell of Ref.~\cite{chen2020development} and the LFP/graphite cell of Ref.~\cite{prada2013simplified}) at a nominal $1$C operating point, and an $\alpha$-MnO$_2$ intercalation pseudocapacitor~\cite{guillemet2012modeling} at a higher-rate ($10$C) point; the arrow shows the direction of increasing C-rate ($Da_p \propto 1/\text{C-rate}$ at fixed $Da_w$). \emph{Right:} the $(Da_p, Da)$ plane measuring electrolyte limitation, with the region where the electrolyte is not rate-limiting ($Da \lesssim Da_p$) shaded; at $1$C the battery chemistries lie below the diagonal, while the $10$C MnO$_2$ pseudocapacitor sits just above it at the onset of electrolyte limitation. The Li-ion parameters are taken from the cited parameterizations, with representative $1$\,M LiPF$_6$ carbonate-electrolyte transport from Ref.~\cite{bernardi2011analysis} where a study does not report its own.}
    \label{fig:envelope}
\end{figure}

The lean model derived in Section~\ref{sec:approach} can be used to map electrochemical storage systems. Figure~\ref{fig:envelope} (left) locates six electrodes drawn from published parameterizations in the $(Da_p,\,Da_w)$ plane. Three are Li-ion positive electrodes: a micron-particle NMC811 electrode~\cite{chen2020development}, a nano-LFP electrode~\cite{prada2013simplified}, and a LTO electrode~\cite{kashkooli2016nano}. Two are graphite negative electrodes, one from the same LG M50 cell as the NMC811~\cite{chen2020development} and one from the LFP/graphite cell of Ref.~\cite{prada2013simplified}. The sixth is an $\alpha$-MnO$_2$ intercalation pseudocapacitor~\cite{guillemet2012modeling}. The Li-ion electrodes use representative $1$\,M LiPF$_6$ carbonate-electrolyte transport from Ref.~\cite{bernardi2011analysis} where a study does not report its own; each is placed at a nominal $1$C operating point, except the pseudocapacitor at a higher rate ($10$C). The exchange current density is evaluated at half charge from each study's reaction-rate constant (taken directly for MnO$_2$, LFP, and the Prada graphite, which report it as an exchange current density), and the specific area from its particle radius and active fraction. Because $Da_p \propto t_p \propto 1/\text{C-rate}$ while $Da_w$ is rate-independent, changing the C-rate slides a device horizontally along the indicated ``operating line'' (arrow), with higher rates to the left; a single marker therefore fixes each electrode at its stated rate. All six lie within the lean-model validity region $Da_w \lesssim 10^2\,Da_p$. The two graphite anodes occupy the low-$Da_p$, low-$Da_w$ corner of the set, because their moderate exchange current and specific area keep reaction comparable to transport and the C-rate; the thinner Prada coating gives the smaller $Da_w$ of the two. The micron-particle NMC811 and the nano-LFP electrodes cluster just above them at intermediate $Da_p$ and $Da_w$ of order unity: nano-LFP has a very high specific area, but its low exchange current keeps reaction from outpacing transport. The LTO electrode, by contrast, combines high specific area with fast kinetics and sits far to the upper right at the largest $Da_p$ and $Da_w$ of the set. The MnO$_2$ pseudocapacitor has the largest wiring-to-process ratio $Da_w/Da_p$ of the set, a consequence of its high specific area and higher operating rate, placing it closest to the validity boundary, though still comfortably within it.

Figure~\ref{fig:envelope} (right) shows the same electrodes in the $(Da_p,\,Da)$ plane, which measures electrolyte limitation. The ratio of the electrolyte Damk\"ohler number to the process Damk\"ohler number, $Da/Da_p$ (Appendix~\hyperref[sec:apdxa3]{A.3}), measures this limitation, so the diagonal $Da = Da_p$ marks its onset. At $1$C all five Li-ion electrodes sit below the diagonal, the LG~M50 graphite and NMC811 electrodes nearest to it, while the thinner Prada graphite together with the high-area LFP and LTO lie well below the line. Because $Da$ is set by geometry and salt transport alone while $Da_p \propto t_p$, raising the rate slides a device leftward toward the boundary; the MnO$_2$ pseudocapacitor, at $10$C, already sits just above the diagonal at the onset of electrolyte limitation ($Da/Da_p \approx 1.3$), and is driven further into that regime at higher rates. This is the same ratio that enters the discharge-curve characterization of two NMC-111 electrodes in Appendix~\hyperref[sec:apdxa7]{A.7}. 

\subsection{Ragone Plots}

The galvanostatic solution of Section~\ref{sec:approxanalytical} also determines the energy and power densities of the electrode, and hence its Ragone characteristic. At constant current the scaled reaction flux is $\tilde{j} = 1/Da_p$ (fixed by the C-rate), and \emph{retaining} the concentration prefactor of the linearized kinetics $\tilde{j} = -f(\langle\tilde{c}_s\rangle)\,(\Delta\tilde{\phi} - \Delta\tilde{\phi}_{eq})$ gives
\begin{equation}
\Delta\tilde{\phi}_{app}(\langle\tilde{c}_s\rangle) = \Delta\tilde{\phi}_{eq}(\langle\tilde{c}_s\rangle) - \frac{\sqrt{f\,Da_w}\,\coth\sqrt{f\,Da_w}}{f\,Da_p},
\label{eq:vloss}
\end{equation}
with $f = f(\langle\tilde{c}_s\rangle)$. The wiring factor $\sqrt{f\,Da_w}\,\coth\sqrt{f\,Da_w}$ interpolates between a well-wired electrode (factor $\to 1$, purely kinetic loss $1/(f\,Da_p)$) and a wiring-limited one (factor $\to \sqrt{f\,Da_w}$), with the electrolyte group entering through Eq.~(\ref{eq:sigma}). Rather than approximate the concentration prefactor, we evaluate it \emph{exactly} from the linearized ECIT kinetics of Appendix~\hyperref[sec:apdxa6]{A.6}. Because $Da_p \propto 1/\text{C-rate}$, the loss grows in proportion to the applied current, like an ohmic internal resistance whose magnitude is set by $Da_w$ and amplified by $1/f$ as charge accumulates.

Over a discharge the electrode delivers a volumetric energy density $\mathcal{E} = \int V\,dQ$ and an average power density $P = \mathcal{E}/\Delta t$, where $\Delta t$ is the discharge time. The charge passed is $Q = Q_{\max}\langle\tilde{c}_s\rangle$ with capacity $Q_{\max} = \varepsilon_{am} c_{s,\max} F$, and the discharge runs from the charged state $\tilde{c}_{s,0}$ until either the accessible filling window is exhausted or the voltage falls to the cutoff $\tilde{V}_{\text{cut}} = V_{\text{cut}}/(R_gT/F)$, whichever occurs first, at the filling $\tilde{c}_{s,\text{cut}}$. Keeping potentials in units of $R_gT/F$ as elsewhere, the energy and power densities are
\begin{align}
\frac{\mathcal{E}}{Q_{\max}\,R_gT/F} &= \int_{\tilde{c}_{s,0}}^{\tilde{c}_{s,\text{cut}}} \Delta\tilde{\phi}_{app}(\langle\tilde{c}_s\rangle)\, d\langle\tilde{c}_s\rangle, \label{eq:Ehat_general}\\
\frac{P}{j_0 a_p\,R_gT/F} &= \frac{1}{Da_p\,(\tilde{c}_{s,\text{cut}} - \tilde{c}_{s,0})}\,\frac{\mathcal{E}}{Q_{\max}\,R_gT/F}, \label{eq:Phat_general}
\end{align}
since the discharge time is $\Delta t = (\tilde{c}_{s,\text{cut}} - \tilde{c}_{s,0})\,t_p$. Two effects keep the delivered energy below a naive capacity-times-voltage estimate: the energy is the integral of voltage over charge, and the accessible window $\tilde{c}_{s,\text{cut}} - \tilde{c}_{s,0}$ shrinks as the rate rises and the polarization $\sqrt{f\,Da_w}\,\coth\sqrt{f\,Da_w}/(f\,Da_p)$ drives the voltage to the cutoff sooner. Equations~(\ref{eq:vloss})--(\ref{eq:Phat_general}) give the Ragone characteristic entirely in terms of the dimensionless groups, with $Da_p$ setting the rate, $Da_w$ the ``wiring" resistance, and the electrode thermodynamics and kinetics entering only through $\Delta\tilde{\phi}_{eq}$, the prefactor $f$, and the cutoff.

\begin{figure}[H]
    \centering
    \includegraphics[width=\textwidth]{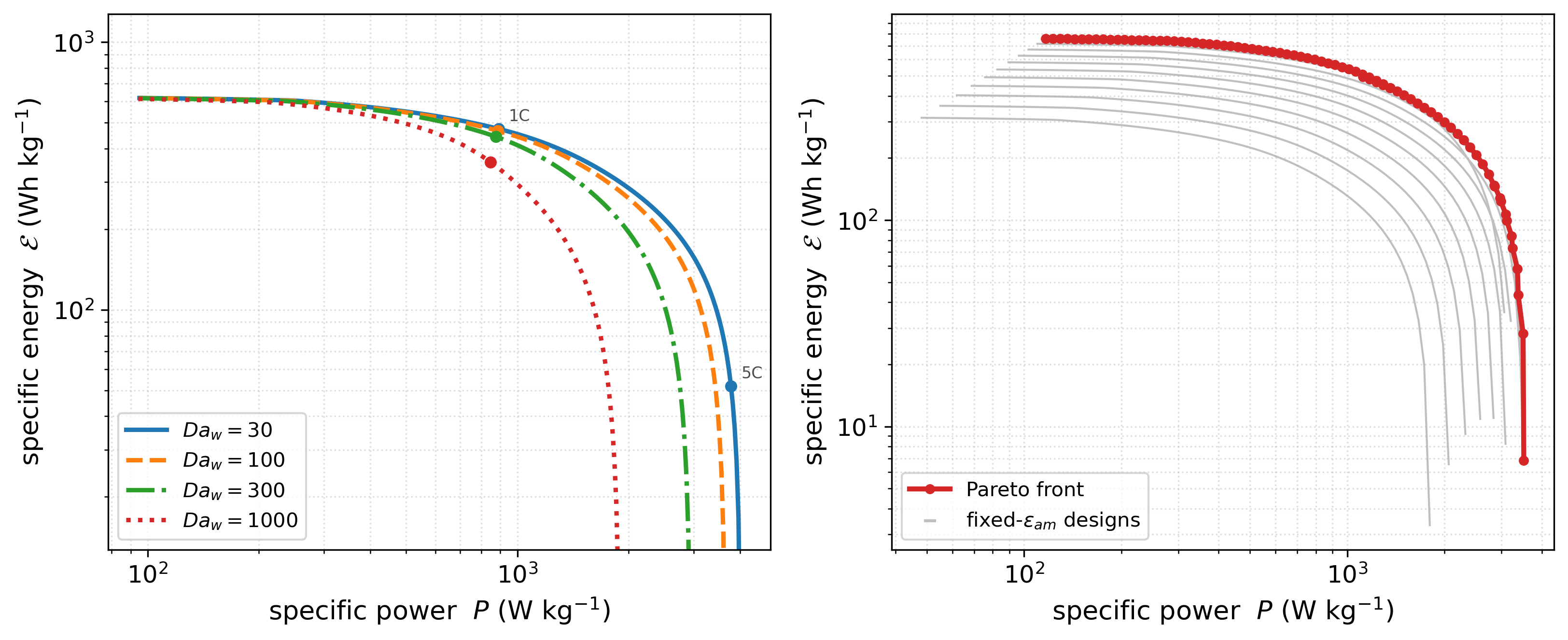}
    \vspace{-0.2cm}
    \caption{Ragone front from the lean model, Eqs.~(\ref{eq:vloss})--(\ref{eq:Phat_general}), using the NMC532 equilibrium voltage~\cite{colclasure2020electrode} and the exact ECIT exchange-current prefactor $f(\langle\tilde{c}_s\rangle)$. Energy and power are specific (gravimetric) quantities for the half-cell working electrode (Wh\,kg$^{-1}$ and W\,kg$^{-1}$). \emph{Left:} single-electrode front for several wiring Damk\"ohler numbers $Da_w$ (through the wiring factor), parametrized by C-rate (markers at $1$ and $5$ C); a larger $Da_w$ moves the high-power knee to lower power. \emph{Right:} design Pareto front: each faint curve is a fixed-$\varepsilon_{am}$ electrode swept over C-rate, with porosity-dependent transport from the Bruggeman relation, and the red envelope is the achievable boundary.}
    \label{fig:ragone}
\end{figure}

Figure~\ref{fig:ragone} (left) evaluates these relations for the NMC half-cell, using the same NMC532 equilibrium voltage~\cite{colclasure2020electrode} as the benchmark simulations and discharging over the filling window $\tilde{c}_{s,0} = 0.30$ to $0.95$. The dimensionless densities are converted to specific (gravimetric) energy and power (Wh\,kg$^{-1}$ and W\,kg$^{-1}$) using the scales $Q_{\max}R_gT/F$ and $j_0 a_p R_gT/F$ of the NMC half-cell working electrode together with an electrode compaction density of $3.7$\,g\,cm$^{-3}$. Here $j_0$ is the ECIT quantum pre-factor of Table~\ref{tab:paramkin}. Plotted in the conventional way, power density against energy density on logarithmic axes, parametrized by C-rate, the curve is nearly flat at low rate, where the deliverable capacity is set by the filling window, and bends down sharply at high rate once the polarization truncates the discharge. A larger $Da_w$ shifts this knee to lower power, so the wiring group governs the rate capability of the electrode.

The trade-off \emph{across designs} follows from the porosity dependence of these scales. Because $Q_{\max} \propto \varepsilon_{am}$ and $j_0 a_p \propto \varepsilon_{am}$, while the wiring factor depends on porosity through the Bruggeman relations $D_{\text{eff}} = \epsilon_p^{\beta} D_0$, $\kappa_{\text{eff}} = \epsilon_p^{\beta} \kappa_0$, and $\sigma_{\text{eff}} = \varepsilon_{am}^{\beta} \sigma_0$ (with $\beta \approx 1.5$; equivalently a tortuosity $\tau = \epsilon_p^{-0.5}$ as in Appendix~\hyperref[sec:apdxa6]{A.6}), each choice of active fraction $\varepsilon_{am} = 1-\epsilon_p$  (assuming no filler fraction) traces its own Ragone curve through Eqs.~(\ref{eq:vloss})--(\ref{eq:Phat_general}). Denser electrodes (larger $\varepsilon_{am}$) store more energy but throttle ionic transport, raising $Da_w$ and the wiring factor; this lowers the attainable power and shrinks the accessible window at high rate. The outer envelope of this family, shown in Figure~\ref{fig:ragone} (right), is the design Pareto front, whose high-power knee reflects the rate-dependent loss of capacity.

\section{Conclusion}
\label{sec:Discuss}
Lithium-ion batteries are complex electrochemical systems whose performance is governed by coupled transport and reaction phenomena across multiple length and time scales \cite{newman1975porous, newman2003modeling}.
Understanding and predicting their behavior through physics-based models has become essential for accelerating cell design, optimizing performance metrics, and enabling advanced battery management strategies \cite{panwar2021recent}.
However, the computational demands of high-fidelity models like the Doyle--Fuller--Newman (DFN) \cite{doyle1993modeling} or Multiphase Porous Electrode Theory (MPET) \cite{smith2017multiphase} frameworks have limited their deployment in applications requiring real-time feedback or large-scale parameter optimization.

This study presents a systematic framework for reaction-limited single-phase electrodes that bridges the gap between computational tractability and predictive accuracy. The main contributions of this work are (i) the derivation of four key dimensionless groups ($Da$, $Da_w$, $Da_p$, $Da_c$) that dictate the electrochemical response of porous electrodes, (ii) practical approximations of PET that yield a ``lean model" of simplified, dimensionless governing equations that capture the essential physics, and (iii) analytical solutions of the lean model under conditions relevant for  batteries. By exploiting simplifications in the reaction-limited regime and adopting a linearized kinetic description motivated by experimental data, we obtained explicit expressions that reveal the quantitative influence of these dimensionless groups on battery response under canonical battery operating protocols (discharge, pulsing, and EIS). Validation against full numerical simulations demonstrated quantitative agreement within practical operating regimes: voltage deviations remained below 60~mV for moderate C-rates ($\leq 1$C), chronoamperometry values matched within $\sim90$\%, and impedance predictions overlapped across the tested parameter space.
This analytical framework enables rapid evaluation of electrode performance without sacrificing the physical basis of porous electrode theory.

Another finding of this work is the identification of limits of the approximation given electrode design and operating conditions.
The approximations perform well when the wiring and process Damköhler numbers satisfy $Da_w \lesssim 10^2 Da_p$, typically corresponding to thin to moderate ($L \lesssim 100~\mu m$) thickness electrodes, good electronic conductivity, and moderate discharge rates.
Beyond these regimes, particularly for thick poorly conducting electrodes or high C-rates where electrolyte polarization becomes significant, corrections accounting for concentration gradients are necessary.
We outlined perturbation-based extensions in Appendix~\hyperref[sec:apdxa3]{A.3} to address these limitations to leading order.

Our framework provides a convenient foundation for the development of reduced-order electrochemical models that retain physical interpretability while achieving computational speeds compatible with embedded systems.
The explicit dependence of performance metrics on dimensionless groups improves interpretability and facilitates sensitivity analysis \cite{li2020parameter}, parameter identifiability studies \cite{bizeray2018identifiability,choi2022parameter, berliner2021nonlinear}, and design space exploration \cite{finegan2021application, xu2021guiding} with limited experimental data.
Further extension to a broader range of electrode materials and chemistries may be possible.
Extension of the linearized kinetics approximation to phase-separating materials such as lithium iron phosphate (LFP) or graphite, where concentration heterogeneities and mosaic instabilities play a central role \cite{zhao2023learning, qi2010situ}, presents a promising avenue for future work on approximations of multiphase porous electrode theory (MPET)~\cite{smith2017multiphase}.
Integration with data-driven approaches, such as hybrid physics-machine learning (hybrid ML) models, could also leverage the computational efficiency of the analytical framework for state estimation and real-time diagnostics in battery management systems \cite{aykol2021perspective, li2022battery, tu2023integrating, borah2024synergizing}.

In summary, the lean model and analytical solutions developed in this work provide an interpretable, quantitatively predictive and computationally light alternative to full numerical simulations for reaction-limited single-phase electrodes.
By making explicit the connection between electrode design parameters and performance, this framework can accelerate battery innovation and enable physics-guided control strategies in next-generation energy storage and ion separation systems.

\section{Acknowledgments}
This research was supported by Shell International Exploration and Production Inc. Additional support was provided by a MathWorks Fellowship from the School of Engineering at MIT (SP). The authors are grateful to Prof. Yang Shao-Horn, Ryan M. Stephens, Harsh Aggarwal, Shreyanil Bhuyan, Yash Samantaray, Armando R. C. Neto for useful discussions, and particularly Shreyanil Bhuyan and Xiaomian Yang for proofreading the equations.

\section*{Appendix}
\subsection*{A1. Derivation of Lean Model} \label{sec:apdxa1}
Here we provide details of the derivation of the simplified PET equations of the lean model analyzed in the main text.
Combining all the species and charge balance equations in Section \ref{sec:model}, the full system of governing equations are,
\begin{align}
    \frac{\partial {c}_s}{\partial {t}} &= \frac{j}{F} \frac{A_p}{V_p} \\
    \epsilon_p \frac{\partial c_l}{\partial t} &= \frac{\partial}{\partial x}\left(\epsilon_p D_{\text{eff}} \frac{\partial c_l}{\partial x}\right) - \frac{1-t_+}{F} a_p j \\ 
    -\frac{\partial}{\partial x}\left( -\kappa_l \frac{\partial \phi_l}{\partial x} + \kappa_l\frac{2R_gT}{F}(1-t_+)\frac{\partial \ln c_l}{\partial x}\right) &= a_p j \\
    -\frac{\partial}{\partial x} \left(-\sigma_s \frac{\partial \phi_s}{\partial x}\right) &= -a_p j
\end{align}
where $j = -j_0f(\tilde{c}_s,\tilde{c}_l)(\Delta \tilde{\phi} -\Delta \tilde{\phi}_{eq})$ is the linearized form of reaction kinetics with a 
general non-linear concentration dependent pre-factor $f(\tilde{c}_s,\tilde{c}_l)$. This is the functional form of kinetics in Section \ref{sec:ctk}.

The scaled species balance in the reaction-limited particles are derived as,
\begin{align}
      \frac{\partial \tilde{c}_s}{\partial \tilde{t}} &= \frac{a_p j_0 t_p}{\epsilon_{am}Fc_{s,max}} \tilde{j} \\ 
      \implies \frac{\partial \tilde{c}_s}{\partial \tilde{t}} &= Da_p \tilde{j} \label{eqn:cseq}
\end{align}
where the concentration scale is $c_{s,max}$, some process timescale $t_p$ and charge transfer current is scaled by $j_0$.

Next, the scaled electrolyte species balance equation in the liquid phase is derived as,
\begin{align}
\frac{\partial^2 \tilde{c}_l}{\partial \tilde{x}^2} &=  \tilde{\tau}_l\frac{\partial \tilde{c}_l}{\partial \tilde{t}} + \frac{j_0 L^2}{F \epsilon_p D_{\text{eff}} c_l^0}(1-t_+) a_p \left(\tilde{j} - \frac{C_{DL} \phi_{\text{ref}}}{j_0 t_p}\frac{\partial \Delta\tilde{\phi}}{\partial \tilde{t}}\right) \\
   \implies \frac{\partial^2 \tilde{c}_l}{\partial \tilde{x}^2} &= \tilde{\tau}_l\frac{\partial \tilde{c}_l}{\partial \tilde{t}} + Da \left(\tilde{j} - Da_c^{-1} \frac{\partial \Delta\tilde{\phi}}{\partial \tilde{t}}\right) \label{eqn:cleq}
\end{align}
where electrolyte concentration scale is $c_l^0$, the voltage scale is $\phi_{\text{ref}}$ and length scale is electrode thickness $L$. $\tilde{\tau}_L = L^2/D_{\text{eff}}t_p$ is the timescale of Li-ion diffusion in the electrolyte scaled by the process time.
Note that an additional capacitive term was introduced to account for any double layer effect which may be significant in related systems like supercapacitors.

Finally, the scaled charge balance equations in the conducting phase and electrolyte phases are derived.
For the solid phase,
\begin{align}
       \epsilon_{am}\sigma_e \frac{\partial^2 \phi_s}{\partial x^2} &=  -a_p \left(j + C_{DL}\frac{\partial (\phi_l - \phi_s)}{\partial t} \right) \\
       \implies \frac{\partial^2 \tilde{\phi}_s}{\partial \tilde{x}^2} &= -\frac{L^2 a_p j_0}{\phi_{\text{ref}} \epsilon_{am} \sigma_e } \left(\tilde{j} - Da_c^{-1} \frac{\partial \Delta \tilde{\phi}}{\partial \tilde{t}}\right)  \label{eqn:phiebal}
       \end{align}
where $\sigma_e$ is the nominal effective conductivity of the conductive backbone.
For the ionically conducting liquid phase,
\begin{align} 
   \frac{2(1-t_+)R_gT \kappa_l}{F \phi_{\text{ref}}} \frac{\partial }{\partial \tilde{x}}\left(\frac{1}{\tilde{c}_l}\frac{\partial \tilde{c}_l}{\partial \tilde{x}}\right) - \kappa_l \frac{\partial^2 \tilde{\phi}_l}{\partial \tilde{x}^2} &= -\frac{L^2 a_p j_0}{\phi_{\text{ref}} \epsilon_p } \left(\tilde{j} - Da_c^{-1}  \frac{\partial \Delta \tilde{\phi}}{\partial \tilde{t}} \right) \label{eqn:cpreapprx}
   \end{align}
We now assume up to mild concentration polarization which makes the following approximation reasonable: $\frac{\kappa_l}{c_l}\frac{\partial c_l}{\partial x} \approx \kappa_l \frac{\partial \tilde{c}_l}{\partial x}$ (this would become exact in the limit of an ideal dilute electrolyte).

Plugging this approximation into Equation \ref{eqn:cpreapprx} gives,
   \begin{align} 
   \frac{2(1-t_+)R_gT \kappa_l}{F\phi_{\text{ref}}} \frac{\partial^2 \tilde{c}_l}{\partial \tilde{x}^2} - \kappa_l \frac{\partial^2 \phi_l}{\partial \tilde{x}^2} &\approx -\frac{L^2 a_p j_0}{\phi_{\text{ref}} \epsilon_p } \left(\tilde{j} - Da_c^{-1}  \frac{\partial \Delta \tilde{\phi}}{\partial \tilde{t}} \right) \\
    \implies \frac{\partial^2 \tilde{\phi}_l}{\partial \tilde{x}^2} &= \left(\frac{L^2 a_p j_0}{\phi_{\text{ref}} \kappa_l \epsilon_p} + \frac{2R_g T (1-t_+) Da}{F \phi_{\text{ref}} } \right)\left(\tilde{j}-Da_c^{-1}  \frac{\partial \Delta \tilde{\phi}}{\partial \tilde{t}}\right)  \label{eqn:philbal}
    \end{align}

Subtracting dimensionless equations \ref{eqn:philbal} and \ref{eqn:phiebal} gives the combined equation for $\Delta \phi$,
\begin{align}
    \implies \frac{\partial^2 \Delta\tilde{\phi}}{\partial \tilde{x}^2} &= -Da_w \left(\tilde{j} - Da_c^{-1} \frac{\partial \Delta\tilde{\phi}}{\partial \tilde{t}}\right) \label{eqn:deltaphieq}
\end{align}

It is instructive to compare the characteristic time constants of the system. The double-layers in Li-ion batteries are thin making the associated charging time $\tau_c = C_{DL}(R_gT/F)/j_0$ fast $\sim 1$s. Furthermore, the electrolyte diffusion time $\tau_l = L^2/D_\text{eff}$ for moderately thin ($100\;\mu m$) electrodes and typical electrolytes is also fast $\sim 10$ s. Therefore, the accumulation terms associated with double-layer charging and electrolyte transport may be neglected to leading order in moderate discharge and pulsing when the process timescale $t_p \sim 1000$s is much slower. Equations \ref{eqn:cseq}, \ref{eqn:cleq} and 
\ref{eqn:deltaphieq} together with the assumptions form the system of scaled equations analyzed in the main text with $\phi_{\text{ref}}=k_BT/e=R_gT/F$.
 
\subsection*{A2.
Solid Solution Approximation} \label{sec:apdxa2}

Consider the following partitioning of the concentration ($\tilde{c}_s$) as $\tilde{c}_s = \langle \tilde{c}_s \rangle + \delta \tilde{c}_s $.
The governing equations then become,

\begin{align}
    \frac{\partial \langle \tilde{c}_s \rangle}{\partial \tilde{t}} + \frac{\partial \delta \tilde{c}_s}{\partial \tilde{t}} = -Da_pf(\tilde{c}_s)(\tilde{\eta} + \Delta \tilde{\phi}_{eq}(\langle \tilde{c}_s \rangle) - \Delta \tilde{\phi}_{eq}) \\ 
    \frac{\partial^2 \tilde{\eta}}{\partial \tilde{x}^2} = Da_w f(\tilde{c}_s)(\tilde{\eta} + \Delta \tilde{\phi}_{eq}(\langle \tilde{c}_s\rangle)- \Delta \tilde{\phi}_{eq})
\end{align}

Expanding the concentration dependent functions about $\langle \tilde{c}_s \rangle $ yields,

\begin{align}
     \frac{\partial \langle \tilde{c}_s \rangle}{\partial \tilde{t}} + \frac{\partial \delta \tilde{c}_s}{\partial \tilde{t}} = - Da_p \overline{f}\tilde{\eta} - Da_p\left(\frac{\partial f}{\partial \tilde{c}_s}\Big|_{\tilde{c}_s=\langle \tilde{c}_s \rangle} \tilde{\eta} - \overline{f} \frac{\partial \Delta\tilde{\phi}_{eq}}{\partial \tilde{c}_s}\Big|_{\tilde{c}_s=\langle \tilde{c}_s \rangle}\right)\delta \tilde{c}_s \\
     \frac{\partial^2 \tilde{\eta}}{\partial \tilde{x}^2} = Da_w \overline{f} \tilde{\eta} 
+ Da_w\left(\frac{\partial f}{\partial \tilde{c}_s}\Big|_{\tilde{c}_s=\langle \tilde{c}_s \rangle} \tilde{\eta} - \overline{f} \frac{\partial \Delta\tilde{\phi}_{eq}}{\partial \tilde{c}_s}\Big|_{\tilde{c}_s=\langle \tilde{c}_s \rangle}\right)\delta \tilde{c}_s
\end{align}
where $\overline{f} = f(\langle \tilde{c}_s \rangle)$.
Spatial averaging of the first and second equations yield,
\begin{align}
    \frac{\partial \langle \tilde{c}_s \rangle} {\partial\tilde{t}} = -Da_p \overline{f}\langle \tilde{\eta}\rangle - Da_p \frac{\partial f}{\partial \tilde{c}_s}\Big|_{\tilde{c}_s=\langle \tilde{c}_s \rangle} \langle \tilde{\eta} \delta \tilde{c}_s \rangle \\ 
    -\frac{\partial \tilde{\eta}}{\partial \tilde{x}}\Big|_{\tilde{x}=0} = Da_w \overline{f} \langle \tilde{\eta} \rangle + Da_w \frac{\partial f}{\partial \tilde{c}_s}\Big|_{\tilde{c}_s=\langle \tilde{c}_s \rangle} \langle \tilde{\eta} \delta \tilde{c}_s \rangle
\end{align}
If the leading order solution for $\tilde{\eta}$ is computed such that one of the boundary conditions is  $\frac{\partial \langle \tilde{c}_s \rangle}{\partial \tilde{t}} = -Da_p \overline{f} \langle \tilde{\eta} \rangle$,  the following relations hold (to leading order),
\begin{align}
    \langle \tilde{\eta} \delta 
\tilde{c}_s \rangle &= 0 \\
    \frac{\partial \delta \tilde{c}_s}{\partial \tilde{t}} &= - Da_p \overline{f}(\tilde{\eta}-\langle \tilde{\eta} \rangle) - Da_p\left(\frac{\partial f}{\partial \tilde{c}_s}\Big|_{\tilde{c}_s=\langle \tilde{c}_s \rangle} \tilde{\eta} - \overline{f} \frac{\partial \Delta\tilde{\phi}_{eq}}{\partial \tilde{c}_s}\Big|_{\tilde{c}_s=\langle \tilde{c}_s \rangle}\right)\delta \tilde{c}_s 
\end{align}

If the timescale of the chosen process is such that $\frac{\partial \tilde{c}_s}{\partial t} \sim Da_p \overline{f}\langle \tilde{\eta} \rangle \sim 1$ (e.g., galvanostatic (dis)charge), the above equation yields the following scaling,

\begin{align}
    \frac{\partial \delta{\tilde{c}_s}}{\partial \tilde{t}} \sim \left(\frac{\tilde{\eta}}{\langle \tilde{\eta} \rangle} -1\right) +\left(\frac{1}{\overline{f}}\frac{\partial f}{\partial \tilde{c}_s}\frac{\tilde{\eta}}{\langle\tilde{\eta}\rangle} + Da_p \overline{f} \frac{\partial \Delta \tilde{\phi}_{eq}}{\partial \tilde{c}_s}\right) \delta \tilde{c}_s
\end{align}
where we have taken $\frac{\partial \langle \tilde{c}_s \rangle}{\partial \tilde{t}} = -Da_p\overline{f}\langle \tilde{\eta}\rangle\sim 1$.
Comparing terms on RHS in the equation above yields a scaling estimate for the magnitude of $\delta \tilde{c}_s$,
\begin{align}
\delta \tilde{c}_s &\sim \frac{1-\frac{\tilde{\eta}}{\langle \tilde{\eta} \rangle} }{ \frac{1}{\overline{f}}\frac{\partial f}{\partial \tilde{c}_s}\frac{\tilde{\eta}}{\langle\tilde{\eta}\rangle} + Da_p \overline{f} \frac{\partial \Delta \tilde{\phi}_{eq}}{\partial \tilde{c}_s}} 
% \tilde{t}_{\text{relax}} &\sim \frac{1}{Da_p \overline{f} \frac{\Delta \tilde{\phi}_{eq}}{\partial \tilde{c}_s}} 
\end{align}
Using the leading order solution for small $Da_w$, $\frac{\tilde{\eta}}{\langle \tilde{\eta} \rangle} = \Lambda\cosh(\Lambda(\tilde{x}-1))/\sinh(\Lambda)$, the condition at which the approximation breaks down is obtained as,
\begin{align}
    \text{max}_x |\delta \tilde{c}_s |
\sim \frac{1-\Lambda/\tanh\Lambda }{ \frac{1}{\overline{f}}\frac{\partial f}{\partial \tilde{c}_s}\Lambda/\tanh\Lambda + Da_p \overline{f} \frac{\partial \Delta \tilde{\phi}_{eq}}{\partial \tilde{c}_s}} \sim  \frac{\Lambda^2}{\frac{1}{\overline{f}}\frac{\partial f}{\partial \tilde{c}_s} + Da_p \overline{f} \frac{\partial \Delta \tilde{\phi}_{eq}}{\partial \tilde{c}_s}} < 1
\end{align}
where $1-\Lambda/\tanh \Lambda$ was approximated to leading order in $\Lambda$.
Plugging $\Lambda^2 = Da_w \overline{f}$,

\begin{align}
    \Bigg|\frac{Da_w}{\frac{1}{\overline{f}^2}\frac{\partial f}{\partial \tilde{c}_s} + Da_p \frac{\partial \Delta \tilde{\phi}_{eq}}{\partial \tilde{c}_s}}\Bigg|
< 1
\end{align}
for electrodes that behave like solid solution such as NMC532 or LCO, $\frac{\partial \Delta \tilde{\phi}_{eq}}{\partial \tilde{c}_s}$ is typically $O(10^2)$ for typical filling fraction range of 0.2-0.8.
Also, for IT-limited CIET kinetics, $f(\tilde{c}_s) \sim \tilde{c}_s^\alpha(1-\tilde{c}_s) \implies \frac{1}{\overline{f}^2}\frac{\partial f}{\partial \tilde{c}_s} \sim O(10)$.
Therefore, the approximation is reasonable as long as $Da_w \lesssim 10^2Da_p$, which provides an upper bound on the applied current. 

\subsection*{A3.
Electrolyte Polarization} \label{sec:apdxa3}

Ion concentration polarization in the electrolyte becomes significant at high discharge rates, where the Damköhler number $Da$ is no longer negligible compared to $Da_p$.
This section derives the leading-order correction to the base model presented in the main text to account for weak electrolyte polarization.
We begin with the coupled, pseudo-steady governing equations for the electrolyte concentration, $\tilde{c}_l$, and the overpotential, $\tilde{\eta}$:
\begin{align}
\frac{\partial^2 \tilde{c}_l}{\partial \tilde{x}^2} &\approx -Da f(\langle \tilde{c}_s \rangle,\tilde{c}_l)\tilde{\eta} \\
\frac{\partial^2 \tilde{\eta}}{\partial \tilde{x}^2} &= Da_w f(\langle \tilde{c}_s \rangle,\tilde{c}_l)\tilde{\eta}
\end{align}
Combining these two equations yields a direct relationship between the concentration and overpotential gradients:
\begin{equation}
    \frac{1}{Da}\frac{\partial^2 \tilde{c}_l}{\partial \tilde{x}^2} = -\frac{1}{Da_w}\frac{\partial^2 \tilde{\eta}}{\partial \tilde{x}^2}
\end{equation}
Integrating this expression twice with respect to $\tilde{x}$ and applying boundary conditions leads to a simplified linear relationship between the local concentration and overpotential.
For weak polarization, we consider a small deviation, $\delta \tilde{c}_l$, from the reference electrolyte concentration, such that $\tilde{c}_l(\tilde{x},\tilde{t}) = 1 - \delta \tilde{c}_l(\tilde{x},\tilde{t})$.
The concentration dependence of the reaction kinetics, $f(\langle \tilde{c}_s \rangle,\tilde{c}_l)$, can be linearized around $\tilde{c}_l=1$:
\begin{equation}
    f(\langle \tilde{c}_s \rangle,\tilde{c}_l) \approx f(\langle \tilde{c}_s \rangle,1)(1-\alpha\delta \tilde{c}_l)
\end{equation}
where $f(\langle \tilde{c}_s \rangle,1)$ is the pre-factor at the reference concentration, and $\alpha = \frac{1}{f}\frac{\partial f}{\partial \tilde{c}_l}\big|_{\tilde{c}_l=1}$ captures the sensitivity of the reaction rate to electrolyte concentration.
By spatially averaging the combined PDE and relating the local deviation of overpotential $\tilde{\eta}$ to the concentration deviation $\delta \tilde{c}_l$, we obtain:
\begin{align}
\label{eq:dcleta}
\delta\tilde{c}_l \approx \frac{Da}{Da_w}\left(\tilde{\eta} -\langle \tilde{\eta} \rangle   - \tilde{\eta}'|_{\tilde{x}=1}\left(\tilde{x}-1/2\right)\right)
\end{align}
where $\langle \tilde{\eta} \rangle$ is the spatial average of the overpotential.
This allows us to reframe the problem as a single linearized ordinary differential equation for the concentration deviation $\delta \tilde{c}_l$:
\begin{align}
\frac{\partial^2 \delta \tilde{c}_l}{\partial \tilde{x}^2} \approx Daf(\langle \tilde{c}_s \rangle) \left(\langle \tilde{\eta} \rangle + \tilde{\eta}'|_{\tilde{x}=1}\left(\tilde{x}-\frac{1}{2}\right) +\left(-\alpha \tilde{\eta}'|_{\tilde{x}=1}\left(\tilde{x}-\frac{1}{2}\right)-\alpha \langle \tilde{\eta} \rangle + \frac{Da_w}{Da}\right)\delta \tilde{c}_l\right)
\end{align}
This equation can be solved subject to mass conservation, $\langle \delta \tilde{c}_l \rangle =0$, and the no-flux boundary condition at the current collector, $\frac{\partial \delta\tilde{c}_l}{\partial \tilde{x}}\big|_{\tilde{x}=1} = 0$.
While the general solution of the above equation can be analytically found, the boundary conditions cannot be applied without numerically integrating Airy and Scorer functions.
Therefore, to arrive at an approximate analytical form further simplifications become necessary.

A reasonable simplification is to assume that under weak electrolyte polarization $Da$ is small compared to $Da_w$ such that, $\tilde{\eta}'|_{\tilde{x}=1} \sim Da_{w,\sigma}/Da_p \ll Da_w/Da$.

Then, the equation can be analytically solved to obtain,

\begin{align}
    \delta \tilde{c}_l = \frac{(A_2-A_1B)\cosh(\sqrt{B}(\tilde{x}-1))- A_2\cosh(\sqrt{B}\tilde{x})}{\sqrt{B}\sinh(\sqrt{B})} + \left(A_1+A_2\left(\tilde{x}-\frac{1}{2}\right)\right)
\end{align}
where $B = \left(Da_w+\alpha \frac{Da}{Da_pf}\right)f(\langle \tilde{c}_s\rangle)$, $A_1 = \frac{Da}{Da_p B}$ and $A_2 = Da_{w,\sigma}fA_1$.
% Substituting new variables $y = \alpha \delta \tilde{c}_l $, $z = \tilde{x}\sqrt{\alpha Da f}$,  $a = -\tilde{\eta}'|_{\tilde{x}=1}$, $b = -\langle \tilde{\eta}\rangle + \tilde{\eta}'|_{\tilde{x}=1}/2$ and $d = b + \frac{Da_w}{\alpha Da}$ yields a simpler form of the ODE  that is given by,
% \begin{align}
%     \frac{\partial^2 y}{\partial z^2} = -(a\tilde{x}+b) + (a\tilde{x}+d)y
% \end{align}

% This can further be simplified to the form,
% \begin{align}
%     \frac{\partial^2 y}{\partial z^2} = \left(a\frac{z}{\sqrt{\alpha Daf}}+d\right)(y-1) + (d-b)
% \end{align}
% Finally, substituting $Z = \left(\frac{a}{\sqrt{\alpha Daf}}\right)^{-2/3}\left(az/\sqrt{\alpha Daf}+d\right)$ and $Y = (y-1)$ yields,
% \begin{align}
%       \frac{\partial^2 Y}{\partial 
% Z^2} = Z.Y + (d-b)\left(\frac{a}{\sqrt{\alpha Daf}}\right)^{-2/3}
% \end{align}
% This ODE is the non-homogeneous Airy Equation with constant forcing term  and has a known solution that is given by:
% \begin{equation}
% Y = c_1 Ai(Z)+c_2 Bi(Z) + \pi (d-b)\left(\frac{a}{\sqrt{\alpha Daf}}\right)^{-2/3} Hi(Z)
% \label{eqn:dcplus}
% \end{equation}

% where $B = (Da_w - \alpha \langle \tilde{\eta}\rangle Da)f(\langle \tilde{c}_s \rangle)$ and $A = Da f(\langle \tilde{c}_s \rangle) \langle \tilde{\eta} \rangle$.
The final step is to apply the galvanostatic current constraint, $\int_{0}^{1} Da_p \tilde{j} d\tilde{x} = 1$.
Substituting the linearized expressions for $\tilde{j}$ and keeping only leading-order terms yields an expression for the average overpotential:
\begin{equation}
\langle \tilde{\eta} \rangle \approx -\frac{1}{Da_p f(\langle \tilde{c}_s \rangle)} \label{eqn:xiavg}
\end{equation}
Finally, the overpotential, $\tilde{\eta}$ can be found by plugging $\delta \tilde{c}_l$ in Equation \ref{eq:dcleta} to obtain the overpotential $\tilde{\eta}(\tilde{x}) $ as,

\begin{align}
    \tilde{\eta} =  \frac{Da_w}{Da}\frac{(A_2-A_1B)\cosh(\sqrt{B}(\tilde{x}-1))- A_2\cosh(\sqrt{B}\tilde{x})}{\sqrt{B}\sinh(\sqrt{B})}
\end{align}

The above expression for $\tilde{\eta}$ can be substituted in Equation \ref{eq:phisintermsofeta} in the main text to get the applied voltage ($\tilde{\phi}_{s,app}$).
This completes the first-order correction for weak electrolyte polarization, linking the increased voltage drop to the Damköhler numbers ($Da$, $Da_w$, $Da_p$) and the kinetic parameters ($f$, $\alpha$).
\subsection*{A4. Hierarchical Electrodes} \label{sec:hierarch}
Hierarchical effects may be important in polycrystalline cathode microstructures with secondary particles composed of agglomerated primary particles and some internal porosity \cite{xu2020charge}.
In this case, the governing equations are multiscale with (i) electrode and (ii) secondary particle (sp) levels.
This picture is helpful and has been used in prior work \cite{lian2024modeling}.
Consider the agglomerate secondary particle to be spherical with a characteristic size $R$ and an apparent surface current density of $\tilde{j}^{sp}$.
Then the dimensionless system of equations at the electrode level may be re-framed with the scaled current density $\tilde{j}_l$ appropriately replaced by $\tilde{j}_{sp}$.
\begin{align} 
    \frac{\partial^2 \tilde{c}_l}{\partial \tilde{x}^2} = Da \tilde{j}_{sp} \\ 
    \frac{\partial^2 \Delta\tilde{\phi}}{\partial \tilde{x}^2} = -Da_w \tilde{j}_{sp}
\end{align}

$j_{sp}$ is dependent on the transport and charge transfer kinetics within an agglomerate particle.
Define $\tilde{c}^{sp}_l (\tilde{r}=1,\tilde{x},\tilde{t}) = \tilde{c}_l(\tilde{x},\tilde{t})$ at a specific value of $\tilde{x}=\tilde{x}_0$.
Similarly, also define  $\tilde{\Delta \tilde{\phi}}^{sp}_l (\tilde{r}=1,\tilde{x},\tilde{t}) = \Delta \tilde{\phi}(\tilde{x}, \tilde{t})$.
\begin{align} 
     \frac{\partial \tilde{c}_s}{\partial \tilde{t}} =  Da_p^{sp} f(\tilde{c}_s,\tilde{c}_l)(\Delta \tilde{\phi}_{eq} - \Delta\tilde{\phi}^{sp}) \\
   \tilde{\nabla}_r^2\tilde{c}^{sp}_l = Da^{sp} f(\tilde{c}_s,\tilde{c}^{sp}_l) (\Delta \tilde{\phi}_{eq} - \Delta\tilde{\phi}^{sp}) \\ 
   \tilde{\nabla}_r^2 (\Delta \tilde{\phi}^{sp}) = -Da_w^{sp} f(\tilde{c}_s,\tilde{c}^{sp}_l) (\Delta \tilde{\phi}_{eq} - \Delta\tilde{\phi}^{sp}) \label{eqn:phisp}
\end{align}

where $\tilde{\nabla}_r$ is the r-laplacian in dimensionless spherical coordinates scaled by particle size $R$.
The boundary conditions for the governing equations in $\tilde{c}_l^{sp}$ and $\Delta \tilde{\phi}^{sp}$ are symmetry at $\tilde{r}=0$ and a matching Dirichlet boundary condition for each variable at $\tilde{r}=1$.
These are given by $\tilde{c}^{sp}_l (\tilde{r}=1,\tilde{x},\tilde{t}) = \tilde{c}_l(\tilde{x},\tilde{t})$ and  $\tilde{\Delta \tilde{\phi}}^{sp}_l (\tilde{r}=1,\tilde{x},\tilde{t}) = \Delta \tilde{\phi}(\tilde{x}, \tilde{t})$.

The new dimensionless groups at the secondary particle level $Da_p^{sp}$, $Da^{sp}$ and $Da_w^{sp}$ are analogously defined.
\begin{align}
Da_p^{sp} &= \frac{t_p a_{sp} j_0}{(1-\epsilon_{sp}) F c_{s,max}} \\
    Da^{sp} &= \frac{j_0 R^2 a_{sp}(1-t_+)}{F \epsilon_{sp} D_{eff} c_l^0} \\
    Da_w^{sp} &= \frac{Fj_0 R^2 a_{sp}}{R_gT}\left(\frac{1}{\sigma^{sp}_e} + \frac{1}{\kappa^{sp}_l}\right) + 2(1-t_+)\frac{Da^{sp}}{\kappa_l^{sp} } \\ 
\end{align}

\noindent where the characteristic size of the secondary particle ($R$), its porosity ($\epsilon_{sp}$), and the active surface area-to-volume ratio in the secondary particle ($a_{sp}$) are used.
 
The effective electronic ($\sigma_e^{sp}$) and ionic conductivity ($\kappa_l^{sp}$) in the pores of the secondary particle are defined according to effective medium theory \cite{bruggeman1935berechnung}.
For simplicity, assume negligible electrolyte polarization. Then, the solution to governing equation for $\Delta \tilde{\phi}^{sp}$ is readily obtained in terms of spherical Bessel functions,

\begin{align}
    \Delta \tilde{\phi}^{sp}(\tilde{r},\tilde{{x}},\tilde{t})-\Delta \phi_{eq} = (\Delta \tilde{\phi}(\tilde{x},\tilde{t})-\Delta \phi_{eq}) \frac{\sinh(\sqrt{Da_w^{sp}f}\tilde{r})}{\sinh(\sqrt{Da_w^{sp}f})\tilde{r}}
\end{align}

Scaled current density $\tilde{j}_{sp}$ in the spherical secondary particle may be obtained as an integral over the particle,

\begin{align}
    \tilde{j}_{sp} = 3\int_{\tilde{r}=0}^{1}  f(\tilde{c}_s,\tilde{c}_l)(\Delta \phi_{eq}-\Delta \tilde{\phi}^{sp}) \tilde{r}^2 d\tilde{r} \\ 
    \tilde{j}_{sp} = 3f(\tilde{c}_s,\tilde{c}_l) (\Delta \phi_{eq}-\Delta \tilde{\phi}(\tilde{x},\tilde{t})) \int \frac{\sinh(\sqrt{Da_w^{sp}f}\tilde{r})}{\sinh(\sqrt{Da_w^{sp}f})\tilde{r}} \tilde{r}^2 d\tilde{r}
\end{align}
where the factor of 3 outside the integral arises due to spherical geometry.
Plugging this integral back in Equation \ref{eqn:phisp} gives the electrode level governing equation for $\Delta \tilde{\phi}(\tilde{x},\tilde{t})$ as,

\begin{align}
    \frac{\partial^2 \Delta \tilde{\phi}}{\partial \tilde{x}^2} = -3 \frac{Da_w}{Da_w^{sp}} (\sqrt{Da_w^{sp}f(\tilde{c}_s,\tilde{c}_l)}\coth(\sqrt{Da_w^{sp}f(\tilde{c}_s,\tilde{c}_l)}) -1)(\Delta \tilde{\phi}_{eq} - \Delta \tilde{\phi})\label{eqn:pcsol}
\end{align}
\noindent The solution has the same form as Equation \ref{eqn:deltaphieq}, with an effective wiring pre-factor of
 $Da'_w = \frac{3}{f(\tilde{c}_s,\tilde{c}_l)} \frac{Da_w}{Da_w^{sp}} (\sqrt{Da_w^{sp}f(\tilde{c}_s,\tilde{c}_l)}\coth(\sqrt{Da_w^{sp}f(\tilde{c}_s,\tilde{c}_l)}) -1)$.
This pre-factor scales differently with $f(\overline{c}_s,\tilde{c}_l)$ depending on the magnitude of $Da_w^{sp}f(\tilde{c}_s,\tilde{c}_l)$.
$Da'_w \sim Da_w f(\tilde{c}_s,\tilde{c}_l)$ when $Da_w^{sp}f(\tilde{c}_s,\tilde{c}_l) \ll 1$ and $Da'_w \sim \sqrt{Da_w Da_w^{sp}f(\tilde{c}_s,\tilde{c}_l)}$ when $Da_w^{sp}f(\tilde{c}_s,\tilde{c}_l) \gg 1$.
 
\subsection*{A5.
Model Identifiability}
The parameters in a battery model are often fitted to observed data and then used to quantify battery performance and aging. This approach can be especially useful when the parameters are based on underlying physics. Yet, doing this reliably has remained difficult due to limited data availability and over-parameterized model structures. Definitive model-driven decisions require the parameters to be uniquely estimable from real-world data \cite{brady2020quantitative}. In other words, the parameters should be identifiable \cite{cole2020parameter}.
Several identifiability studies of MPET, DFN, and SPM models test this using a variety of methods, such as analyses of loss-function contours, chi-squared error, and sensitivity \cite{berliner2021nonlinear, choi2022parameter,galuppini2023nonlinear, berliner2025bayesian}.

In this section, we test key aspects of identifiability for our model.
Practical identifiability of the model parameters is first tested using a
Markov chain Monte Carlo (MCMC) algorithm~\cite{geyer1992practical}. This
tests whether the model fit to observed data is sufficiently sensitive to
yield unique parameter estimates. Furthermore, since our model is nonlinear,
we also analyze contours of the chi-square estimator to verify the existence
of a well-defined global minimum.

Synthetic 1C discharge data are generated for the analytical model using dimensionless numbers derived from NMC532 electrode parameters. To emulate experimental uncertainty, 5\% multiplicative Gaussian noise is applied independently to the wiring and process Damk\"ohler numbers ($Da_w$ and $Da_p$).

\begin{figure}[H]
    \centering
    \includegraphics[width=\textwidth]{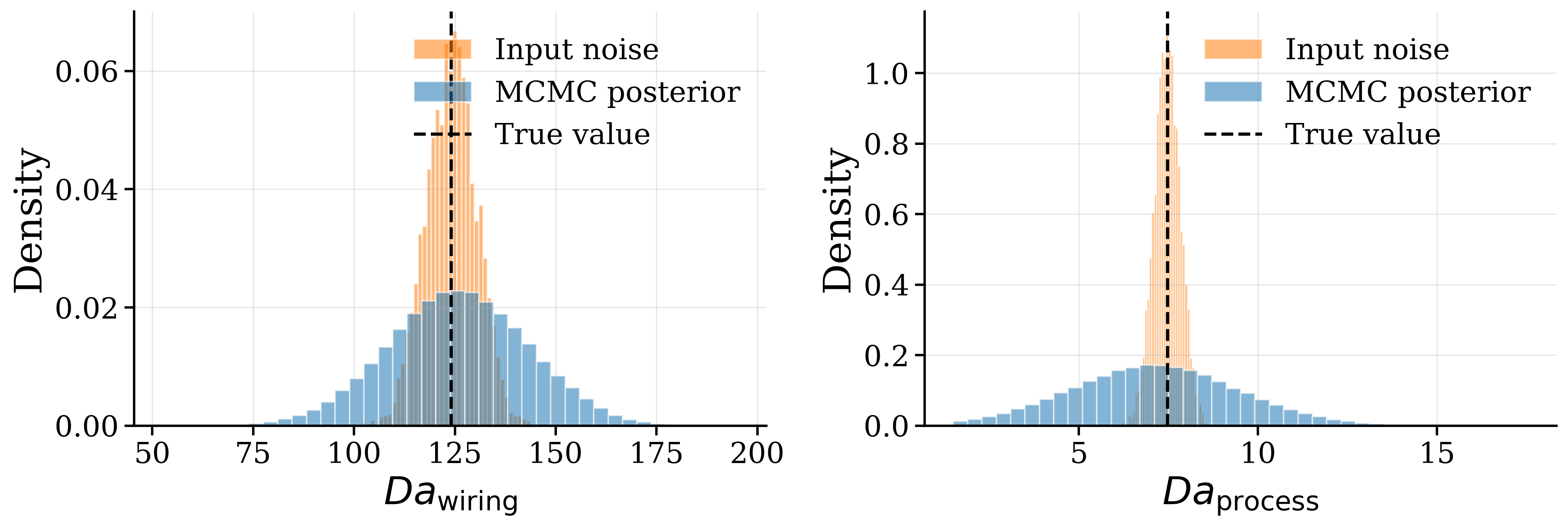}
    \vspace{-0.2cm}
    \caption{Noisy data and posterior MCMC ($Da_{w}, Da_{p}$) inferred from galvanostatic (1 C) model data with Gaussian noise ($\pm$ 5\% s.d.) added to $Da_{w}$ and $Da_{p}$}
    \label{fig:mpetmcmc}
\end{figure}

Figure~\ref{fig:mpetmcmc} shows the posterior distributions of $Da_w$ and $Da_p$ obtained via affine-invariant ensemble MCMC sampling~\cite{foreman2013emcee}. Both marginal distributions are
unimodal and centered on the true parameter values, confirming good
practical identifiability. 

\begin{figure}[H]
    \centering
\includegraphics[width=0.75\textwidth]{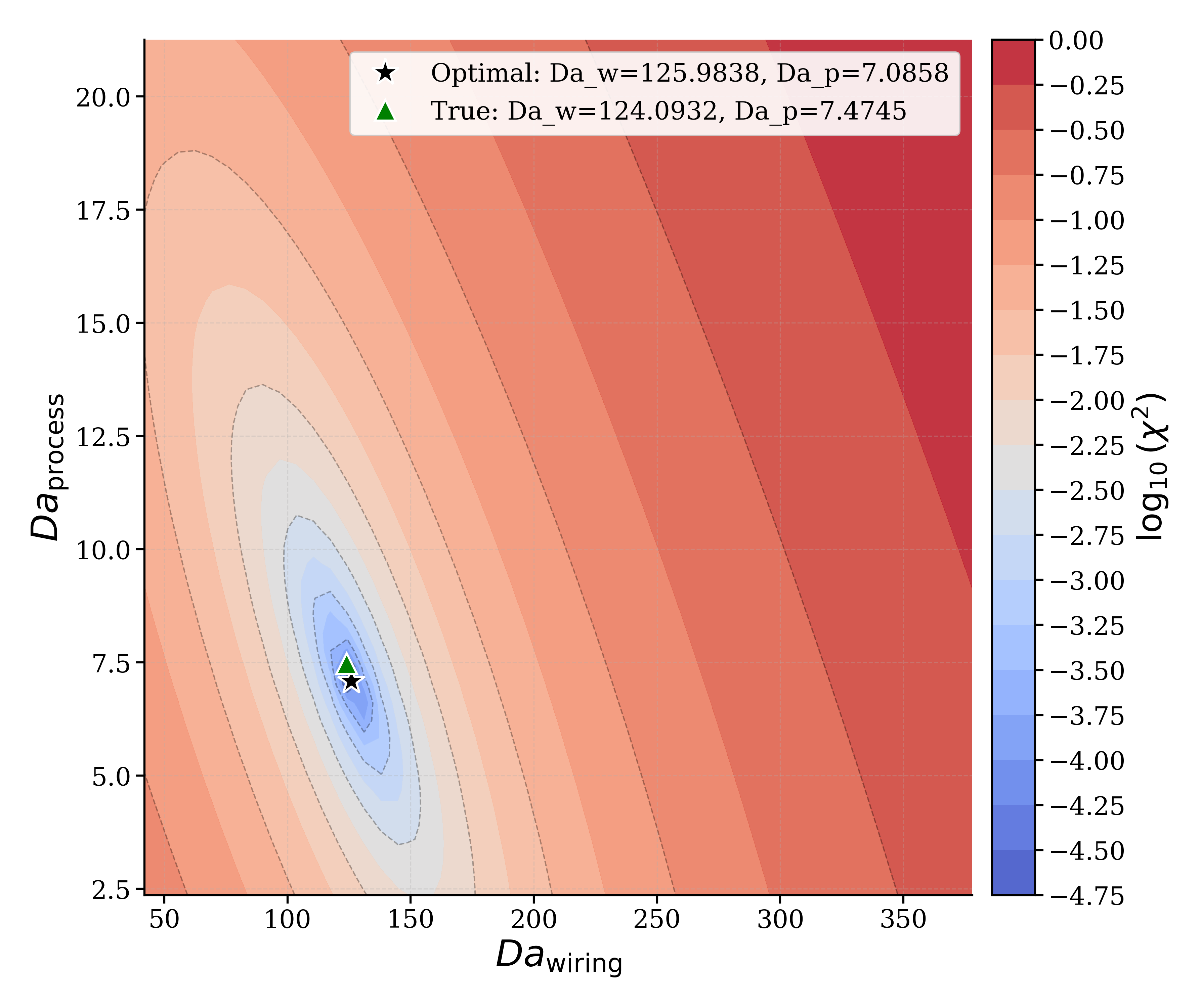}
    \vspace{-0.2cm}
    \caption{$\chi^2$ landscape for fitting galvanostatic (1 C) model data with Gaussian noise ($\pm$ 5\% s.d.) added to $Da_{w}$ and $Da_{p}$}
    \label{fig:mpetchisquare}
\end{figure}

Complementing the MCMC analysis, Figure~\ref{fig:mpetchisquare}
displays the $\chi^2$ landscape in $(Da_w,\, Da_p)$ space. The landscape
exhibits a single, well-defined global minimum whose location closely
coincides with the true parameter values. The roughly elliptical
contours, elongated along the direction of the $Da_w$--$Da_p$
anti-correlation seen in the MCMC posterior, confirm that one complete galvanostatic
voltage response contains enough information to get a good estimate for both dimensionless groups.

\newpage
\subsection*{A6.
Parameters for NMC Half-Cell} \label{sec:apdxa6}
\begin{table}[h]
\centering
\caption{Base parameters used in benchmark half-cell simulations}
\label{tab:parameters}
\begin{tabular}{lccc}
\toprule
Parameter & Symbol & Value & Unit \\
\midrule
Positive electrode open circuit potential (OCP) & $\phi_{eq}$ & Ref.
\cite{colclasure2020electrode} & V \\
Positive electrode thickness & $L$ & $10^{-4}$ & m \\
Positive electrode porosity & $\epsilon$ & 0.5 & - \\
Positive electrode active material loading fraction & - & 0.69 & - \\
Cathode particle radius & $R_p$ & $5 \times 10^{-7}$ & m \\
Positive electrode bulk conductivity & $\sigma_{s}$ & 0.1 & S\,m$^{-1}$ \\
Cathode kinetics & - & Table 2 & - \\
Maximum concentration (NMC532) & $c_{s,max}$ & $4.95 \times 10^{4}$ & mol\,m$^{-3}$ \\
Separator thickness & $L_{sep}$ & $5 \times 10^{-6}$ & m \\
Separator porosity & $\epsilon_{sep}$ & 1 & - \\
Initial electrolyte concentration & $c_{0}$ & 1000 & mol\,m$^{-3}$ \\
Cation transference number & $t_+$ & 0.38 & - \\
Electrolyte conductivity & $\kappa_l$ & Ref.
\cite{bernardi2011analysis} & S\,m$^{-1}$ \\
Electrolyte diffusivity parameters & $D$ & Ref.
\cite{valoen2005transport} & m$^2$\,s$^{-1}$ \\
Bruggeman exponent (positive electrode) & - & -0.5 & - \\
Reference temperature & $T_{ref}$ & 298.15 & K \\
\bottomrule
\end{tabular}
\end{table}
The baseline benchmark simulations use the ECIT expression (the ET-limited asymptotic form of CIET) \cite{bazant2023unified}.
The expression was introduced in Section \ref{sec:ctk} and is reiterated below,
\begin{align}
    j_{ECIT} &= j_0(1-\tilde{c}_s) \left(\frac{\tilde{c}_s}{1+\exp(-\tilde{\eta}_f)}-\frac{\tilde{c}_l}{1+\exp(\tilde{\eta}_f)}\right)\text{erfc}\left(\frac{\tilde{\lambda}+\sqrt{1+\sqrt{\tilde{\lambda}}+\tilde{\eta}_f^2}}{2\sqrt{\tilde{\lambda}}}\right)
    \label{eq:fecit}
\end{align}
where $j_0$ is the quantum mechanical pre-factor, $\eta_f$ is the formal overpotential, $\lambda$ is the reorganization energy in the electrode material, $\tilde{c}_s$ is the solid filling fraction, $\tilde{c}_l$ is the adsorbed $\text{Li}^+$  concentration at the interface for an electrolyte with $\text{Li}^+$ activity $a_+$ and adsorption free energy $w_{ads}$,
\begin{align}
    \tilde{c}_l = \frac{a_+e^{-\tilde{w}_{ads}}}{1+a_+e^{-\tilde{w}_{ads}}}
\end{align}
\begin{table}[h]
    \centering
    \begin{threeparttable}
        \caption{Parameters for cathode kinetics}
        \label{tab:paramkin}
  
        \begin{tabular}{lccc}
            \toprule
            Parameter & Symbol & Value & Unit \\
            \midrule
            Exchange current pre-factor & $j_0$ & 5\tnote{a} & A\,m$^{-2}$ \\
            Reorganization Energy & $\lambda$ & 0.11\tnote{a} & eV \\ 
         
           Adsorption free energy & $w_{ads}$ & 25\tnote{a} & meV \\
            Electrolyte activity  & $a_+$ & 1.9\tnote{a} & - \\
            \bottomrule
        \end{tabular}
        \begin{tablenotes}
            \small % Optional: Makes the footnote text slightly smaller
            \item[a] Nominal values based on CIET parameters 
for NMC electrode materials in Ref. \cite{zhang2025lithium}
        \end{tablenotes}
    \end{threeparttable}
\end{table}
% where we use the scaling, $-\ln f \sim \Delta \tilde{\phi}_{eq}$ based on the typical power law form of exchange current density in terms of activity of the intercalant $a_R$ in Butler-Volmer kinetics, $f \sim a_R^\alpha$ ($0<\alpha <1$).
%Aside.. speed 
% \begin{table}
%     \caption{\label{tab1}Table Template.}
%     \begin{tabular*}{\textwidth}{@{}l*{15}{@{\extracolsep{0pt plus12pt}}l}}
%     \br
%     Column 1&Column 2&Column 3\\
%     \mr
% \\
%     list 1&list 2&list 3\\
%     \br
%     \end{tabular*}
% \end{table}

\newpage
\subsection*{A7. Demonstration: characterizing NMC-111 in the lean framework} \label{sec:apdxa7}

The scaling analysis above also serves as a compact characterization tool: fitting the leading-order discharge solution to measured rate data returns a small set of dimensionless groups that fingerprint an electrode and locate it among the limiting mechanisms. We demonstrate this on the two LiNi$_{1/3}$Mn$_{1/3}$Co$_{1/3}$O$_2$ (NMC-111) cathodes of Ren et al.~\cite{ren2019ultrahigh}: a high-power material (HP-NMC) whose plate-like crystallites present radially oriented Li-diffusion channels, and a commercial reference (c-NMC) of the same nominal chemistry. That study traces the markedly better rate capability of HP-NMC to faster intra-particle Li transport set by this microstructure rather than to electrolyte or cell-level effects. The lean model offers a quantitative test of that attribution: if the difference is intra-particle, it should appear in the dimensionless groups.

We digitized the galvanostatic half-cell discharge curves at $0.5$C, $1$C, and $2$C for both electrodes and used the low-rate measurement as the open-circuit reference $U(x)$. The galvanostatic voltage follows from the leading-order solution Eq.~(\ref{eq:vloss}), with the wiring group taken in its electrolyte-corrected form $B = f\,Da_w + \alpha\,Da/Da_p$ from Appendix~\hyperref[sec:apdxa3]{A.3} (with $\alpha\approx\tfrac12$) and a lumped series resistance $R_s$ added for the rate-proportional ohmic drop. We report this series resistance as a gravimetric value $R_s$ (in $\Omega$\,g) using the nominal specific capacity of $\sim$175~mAh~g$^{-1}$ from Ref.~\cite{ren2019ultrahigh}, so that the $1$C current is $I_{1\mathrm{C}}=0.175$~A~g$^{-1}$. We keep the exchange-current prefactor in the pure ECIT/CIET form $f(\tilde c_s)=f_{\mathrm{ECIT}}(\tilde c_s)$ of Eq.~(\ref{eq:fecit}). As is standard for porous electrodes, only a fraction of the nominal host capacity is electrochemically accessible; we capture this with a single usable-capacity fraction per electrode that rescales the stoichiometry window so the surface filling saturates ($\tilde c_s\to1$) at the end of the usable window, which also sets the steep end-of-discharge cutoff. For each electrode we fit seven parameters: the start stoichiometry $a$, the usable-capacity fraction, the wiring group $Da_w$, the electrolyte Damk\"ohler number $Da$, the process Damk\"ohler number $Da_p$, and the series resistance $R_s$.

\begin{table}[h]
\centering
\caption{Lean-model descriptors fit to the NMC-111 discharge data of Ren et al.~\cite{ren2019ultrahigh} ($0.5$C, $1$C, $2$C half-cell discharge). The process Damk\"ohler number $Da_p$ is quoted at $1$C; the series resistance is reported as a gravimetric value using the nominal $175$~mAh~g$^{-1}$ capacity ($I_{1\mathrm{C}}=0.175$~A~g$^{-1}$); the RMS is the discharge-curve voltage residual over all three rates.}
\label{tab:ren}
\begin{tabular}{lccc}
\toprule
Descriptor & Symbol & HP-NMC & c-NMC \\
\midrule
Wiring group & $Da_w$ & $1.6\times10^{2}$ & $5.0\times10^{3}$ \\
Electrolyte Damk\"ohler number & $Da$ & $3.8\times10^{3}$ & $4.2\times10^{3}$ \\
Process Damk\"ohler number (at $1$C) & $Da_p$ & $3.5\times10^{2}$ & $2.5\times10^{2}$ \\
Series resistance ($\Omega$\,g) & $R_s$ & 0 & 0.14 \\
\midrule
Discharge-curve RMS (mV) & --- & 49 & 33 \\
\bottomrule
\end{tabular}
\end{table}

\begin{figure}[H]
\centering
\includegraphics[width=\textwidth]{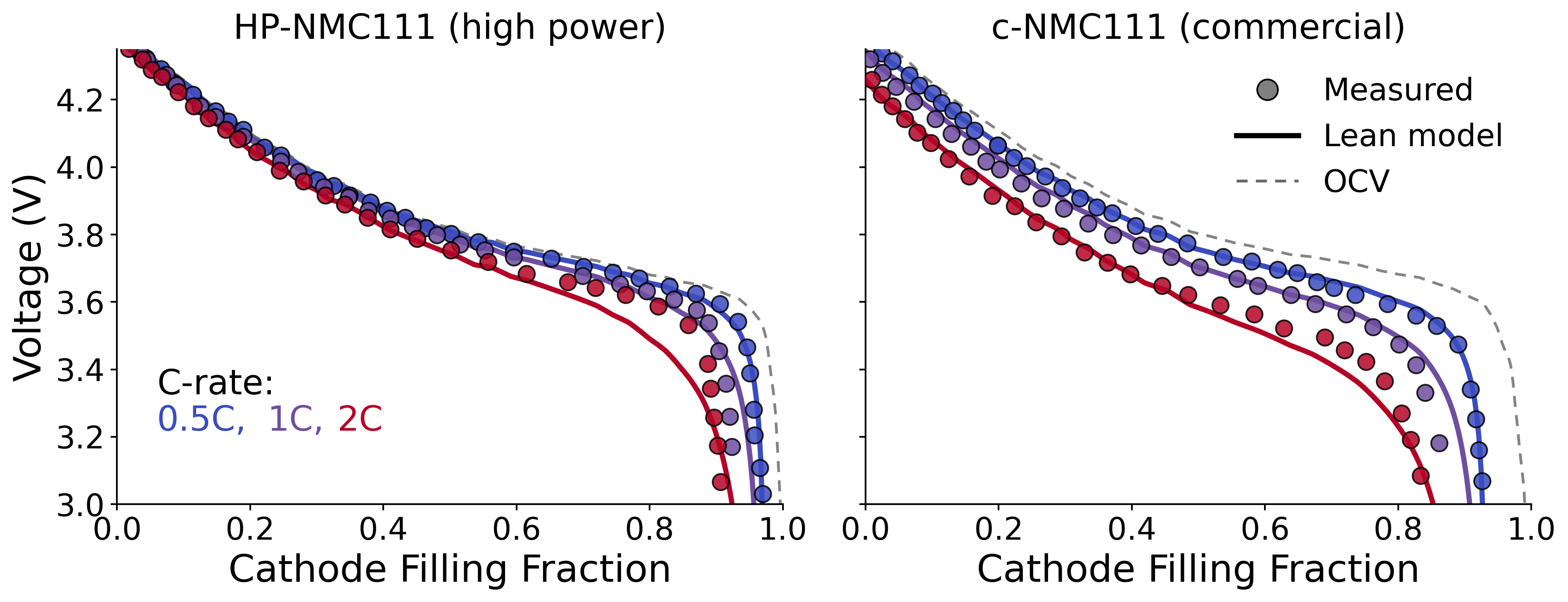}
\caption{Lean-model discharge characterization (Eq.~\ref{eq:vloss}) of the two NMC-111 electrodes of Ren et al.~\cite{ren2019ultrahigh} at $0.5$C, $1$C, and $2$C: high-power HP-NMC (left) and commercial c-NMC (right). Filled markers are the measured half-cell data, solid lines the fitted leading-order solution with the pure ECIT prefactor, and the dashed line the open-circuit reference $\Delta \phi_{eq}(\tilde{c}_s)$. A single fit per electrode (Table~\ref{tab:ren}) captures both materials to $\sim$30--50\,mV; the commercial electrode is distinguished by a larger wiring group, series resistance, and electrolyte Damk\"ohler number, and a smaller usable-capacity fraction.}
\label{fig:leandemo}
\end{figure}

Figure~\ref{fig:leandemo} shows that even with the pure ECIT prefactor the leading-order solution reproduces both electrodes across all three rates to within $\sim$30--50\,mV, including the sloping plateau and the sharp end-of-discharge fall-off. The fitted descriptors (Table~\ref{tab:ren}) cleanly separate the two materials: the commercial electrode has a much larger wiring group ($Da_w\approx5\times10^{3}$ vs.\ $1.6\times10^{2}$), a larger series resistance ($R_s\approx0.14$ vs.\ $0$\,$\Omega$\,g), and a larger electrolyte-to-process Damk\"ohler ratio ($Da/Da_p\approx17$ vs.\ $11$ at $1$C), together with a slightly smaller usable-capacity fraction ($0.94$ vs.\ $0.98$). All four point the performance gap toward intra-particle/contact transport and ohmic resistance rather than the electrolyte, so its reaction current is more strongly polarized and it loses accessible capacity faster with rate. This is consistent with the conclusion Ref.~\cite{ren2019ultrahigh} draws from the microstructure: the engineered radial Li-diffusion channels of HP-NMC mitigate the solid-state bottleneck, which the lean model registers as a smaller effective wiring loss and a larger usable fraction.

We note the scope of the characterization. The lean model resolves the porous-electrode and electrolyte scaling but represents each particle by its volume-averaged filling, so intra-particle solid diffusion is not modeled explicitly. Because we retain the pure ECIT prefactor, the residual $\sim$30--50\,mV misfit (largest in the c-NMC mid-discharge) might be due to deviations from the assumed CIET form; accounting for that deviation, through a measured concentration-dependent prefactor $f(\tilde c_s)$ or an explicit solid-diffusion correction, is the natural route to a tighter fit. 

%\section{Figures}

\newpage

\section*{References}

\bibliographystyle{unsrt} % reference style
\bibliography{refs} % Entries are in the refs.bib file

\end{document}